\documentclass[review]{elsarticle}
\usepackage[latin1]{inputenc}

\usepackage{graphicx}				
\usepackage{subfigure}
\usepackage{amssymb}				
\usepackage{xspace}				
\usepackage{upgreek}				
\usepackage{numcompress}

\biboptions{square, sort&compress}
\journal{Carbon}

\def\MnO{MnO$_2$\xspace}
\newcommand{\MnOx}{MnO$_x$\xspace}
\newcommand{\Mnion}{MnO$_4^-$\xspace}

\newcommand{\NaSO}{Na$_2$SO$_4$\xspace}
\newcommand{\NaMn}{NaMnO$_4$\xspace}
\newcommand{\MnOIon}{MnO$_4$\xspace}

\newcommand{\SO}{SO$_4^{2-}$\xspace}
\newcommand{\N}{N$_2$\xspace}
\newcommand{\CO}{CO$_2$\xspace}
\newcommand{\HO}{H$_2$O\xspace}

\newcommand{\CC}{$^\circ$C\xspace}
\newcommand{\mrm}{\mathrm}

\newcommand{\Sext}{$S_\mathrm{ext}$\xspace}

\newcommand{\SBET}{$S_\mathrm{BET}$\xspace}

\newcommand{\Vmic}{$V_\mathrm{mic}$\xspace}

\newcommand{\Cgrav}{$C_\mathrm{grav}$\xspace}
\newcommand{\dpart}{$d_\mathrm{part}$\xspace}
\newcommand{\dpore}{$d_\mathrm{pore}$\xspace}
\newcommand{\mMnOSext}{m$_{\mathrm{MnO_2}}$/\Sext}
\newcommand{\rhoMnO}{$\rho_{\mathrm{MnO_2}}$}

\renewcommand{\deg}{\xspace$^\circ$C\xspace}

\begin{document} 

\begin{frontmatter}

  \title{Structural and electrochemical properties of \MnO-Carbon based supercapacitor electrodes}

  \author[ZAE]{Christian Weber\corref{cor1}}
  \author[ZAE]{Volker Lorrmann}
  \author[ZAE]{Gudrun Reichenauer} 
  \author[ZAE,EP6]{Jens Pflaum}

  \cortext[cor1]{Corresponding author. E-mail address: cweber@physik.uni-wuerzburg.de}
  \address[ZAE]{Bavarian Center for Applied Energy Research, Am Hubland, 97074 Wuerzburg, Germany }
  \address[EP6]{University of Wuerzburg, Experimental Physics 6, Am Hubland, 97074 Wuerzburg, Germany}

  \begin{abstract}
    We present results of an approach to incorporate redox-active manganese oxide into the 3-dimensional porous structure of carbon xerogels under self-limiting electroless conditions.
    By varying the structure of the carbon backbone, we found that deposition of manganese oxide preferably takes place on the external surface area of the carbon xerogels. From our detailed analysis we conclude that 3-dimensional carbon xerogels with particle and pore sizes ranging from 10 to 20~nm and low manganese oxide precursor concentration combined with long deposition times are beneficial for fast operating pseudocapacitance electrodes with high capacitance and effective use of the redox active component.
  \end{abstract}

\end{frontmatter}

\section{Introduction}

  In times of rising energy costs energy efficiency is the most urgent task to be addressed. A possible strategy is the reuse of energy e.g. generated upon deceleration of various kinds of movements in industrial manufacturing processes or automotive applications. This is the key application for  supercapacitors, storing and releasing electrical energy on the timescale of seconds \cite{Conway1999, Miller2008, Koetz2000, Frackowiak2001}. In order to reach higher energy densities on this timescale, research is extending to faradaic storage processes superimposed to the double layer storage in electrochemical capacitors. Due to the characteristic linear dependence of the faradaic current on the applied voltage, this kind of faradaic storage (e.g. in ruthenium dioxide or certain conducting polymers) is called ``pseudocapacitance'' \cite{Conway1999}.

  In 1999 Lee and Goodenough \cite{Lee1999} described a pseudocapacitive behavior of manganese oxides (\MnOx) in aqueous electrolytes. Since then, research activities on this relatively easy to handle and abundantly available material continuously increased, due to its obvious advantages over other pseudocapacitive materials. For instance, compared to the also promising ruthenium dioxides, \MnOx does not require strong acidic electrolytes, is less toxic and much cheaper in production.

  Since the fast redox-charge storage in \MnOx was identified to take place mainly on the surface of the material, efforts were focusing on optimized structural design of hybrid electrodes \cite{Toupin2004}. Such approaches range from simply mixing carbon as a conductive agent, \MnOx and a binder \cite{Beaudrouet2009,Jacob2009,Reddy2004}, to thin film deposition \cite{Cross2011,wei2007} and templating \cite{Dong2006,Lei2008}. 
  Alternatively, electrolytic deposition  of \MnO into (activated) microporous carbon xerogels, as suggested by several groups, e.g. Cross \textit{et al.} \cite{Cross2011}, will employ the micropores at the surface of the primary particles and therefore increases the total amount of \MnO available for charge storage. A promising approach based on electrodeposition was recently published by Lin \textit{et al.}.\cite{Lin2011}, making them able of incorporating high amounts of \MnO into the electrode. By directly contacting the carbon skeleton of the electrodes, they surmount the contact resistance of \MnO at the current collector-electrode interface and obtained electrodes with very high charging rates.

  Fischer \textit{et al.} \cite{Fischer2007, Fischer2008} chose an approach of electroless deposition of \MnO on a microporous carbon aerogel (CA) from aqueous solution of \NaMn under self-limiting conditions. This was found to be an easily scalable and, in terms of experimental effort, not very demanding method to incorporate \MnO into the well interconnected porous structure of the CA. Compared to simply mixing CA and \MnO, coating of the CA particles by a thin layer of \MnO ensures a better electrical contact between the poorly conducting \MnO and the carbon backbone. 

  Tough they demonstrated in their first study \cite{Fischer2007} they showed that for neutral solutions, MnO$_2$ penetrates deeply into the aerogel structure, the exact location of the oxide remained unrevealed, leaving it open whether \MnO resided in the micropores of the carbon backbone or on the enveloping surface area of the carbon xerogel primary particles (``external surface area'').  Since the location of the active material in the electrode is crucial for its efficient use in charge storage, we adapted Fischer's approach on CA reference systems to further investigate the location of the \MnO deposits.

  Due to their variable structural properties, carbon aerogels form an ideal model system to investigate the influence of structural aspects in electrode design on the resulting performance of the electrodes. The three-dimensionally interconnected microporous skeleton of CAs provides huge surface areas advantageously for both double layer charge storage and \MnO deposition, as well as good electrical conductivity. The well-linked network of interparticular mesopores \cite{IUPAC1984a} ensures excellent electrolyte transport. Depending on the size of the primary particles, their envelope surface area, also denoted as external surface area (\Sext), can be easily varied. For our study, we used carbon xerogels, which have the same structural properties as the aerogels described above. However, in contrast to aerogels, xerogels are synthesized by drying the organic precursor sub-critically, making the process substantially easier to handle.

  Addressing the method of electroless deposition, in a first step we analyze the influence of deposition time and manganese oxide precursor concentration on the deposition of \MnOx on the carbon xerogel backbone and their impact on the resulting electrochemical behavior of the electrodes. Based on this information, we determine a relationship between \MnOx deposition and external surface area of the carbon xerogel by varying the properties of the carbon backbone, i.e. the primary particle size and thereby the external surface area.

\section{Experimental}

  \subsection{Preparation of electrodes}
    Fiber reinforced carbon xerogels (CX) were prepared with different particle sizes yet similar density in order to create a carbon backbone with different external surface area but similar porosity and specific micropore volume. To achieve this goal, we used an accelerated sol-gel route as described in \cite{Wiener2004}. Carbon paper (Sigratex, SGL group) was soaked with the sol, having a molar ratio of resorcinol to catalyst (0.1 N NaCO$_3$) of 500, 3000 and 8000 and yielding CXs with small, medium and large primary particles, respectively. The mass ratio of resorcinol and formaldehyde to total mass (including water) was chosen to be 28 for all samples. The molar ratio of resorcinol to formaldehyde (37~\% aqueous solution) was 1:2. After gelation for 24 hours at 85$^\circ$C, the residual water was exchanged for ethanol before subcritical drying. The samples were dried in ambient atmosphere at room temperature for three days. Subsequently, the sheets of fiber supported organic xerogel were pyrolized under argon atmosphere for 60~min at 850\deg, resulting in about 200~$\upmu$m thick, fiber reinforced carbon xerogel sheets. The mass fraction of the carbon fibers was about 40~\%.

    To incorporate the manganese oxide active material into the carbon structure, a (self-limiting) deposition process, as described by Fischer \textit{et al.} \cite{Fischer2007, Fischer2008}, was employed. Briefly, a carbon xerogel was immersed into an aqueous solution of 0.05~M \NaSO and placed under mild vacuum for 30~min to ensure removal of residual gas and thus complete wetting of the xerogel pores. Subsequently, the solution was exchanged for an aqueous mixture of 0.1~M \NaSO and 0.1~M \NaMn. For the variation in precursor concentration, the amounts were adjusted accordingly. The deposition took place over periods of 5~minutes to 24 hours. No evolution of CO$_2$ was observed. We therefore assume a deposition mechanism as described by Ma et \textit{al.}, in which MnO$_2$ is formed on carbon surfaces without releasing carbon during the process \cite{Ma2006}.
    After deposition, the electrodes were rinsed with distilled water until the rinse was optically clear, then immersed in fresh \HO and placed in vacuum for 30~min. This exchange cycle was repeated three times, ensuring comprehensive removal of residual \NaMn. After drying for 24 hours at 85~$^\circ$C, the mass of the hybrid electrodes (CXMn) was determined.

  \subsection{Structural characterization}
    Surface morphology of the samples was investigated with a Zeiss Ultra plus scanning electron microscope (SEM). The pore structure of the bare CX and the CXMn samples was analyzed via nitrogen gas sorption at 77~K. The measurements were performed with a Micromeritics ASAP~2020 System. The samples were degassed prior to sorption analysis in vacuum at 300\CC for 24~h. 
    The crystalline state of \MnO is known to change with temperature \cite{Huang2008}. This morphological modification seems to leave the microstructure unchanged on length scales probed by gas sorption measurements: No significant changes in the isotherms were observed for hybrid samples degassed at 110\CC, compared to isotherms recorded after heating to 300\CC.\\
    The total surface area was determined using BET (Brunauer-Emmet-Teller) theory \cite{Brunauer1938a}; the external surface area \Sext and micropore volume \Vmic were determined by means of the t-plot method \cite{Lippens1965,Harkins1944a}. The size of the primary particles $d_{\mrm{part}}$ of the bare carbon xerogel was estimated from \Sext and the particle density $\rho_{\mrm{part}}$ assuming spherical particle geometry:
      \begin{equation}
      d_{\mrm{part}} = 2 \frac{3}{S_{\mrm{ext}} \cdot \rho_{\mrm{part}}} \label{eqn:dpart}
      \end{equation}
    with 
      \begin{equation}
      \rho_{\mrm{part}} = \frac{1}{\frac{1}{\rho_\mrm{C}} + V_{\mrm{mic}}} \label{eqn:rhopart},
      \end{equation}
    and a carbon bulk-density of $\rho_\mrm{C} = 2.2$~g~cm$^{-3}$ \cite{Wiener2004}.

    The surface chemistry of representative CX and CXMn samples was investigated via x-ray photoelectron spectroscopy (XPS). The spectra were acquired using a monochromatic Al-K$\alpha$ excitation at 1486.6~eV and an Omicron EA 125 electron analyzer. The spectra are normalized to the C1s peak and standard Shirley background correction has been carried out.

  \subsection{Electrochemical characterization}
    For the electrochemical measurements, a circular electrode ($d$ = 19~mm) was cut and weighted. Electrochemical characterization was performed in a home built sandwich cell, made of Teflon with titanium current collectors. The electrode under investigation is separated by a glass paper from a  carbon xerogel electrode, which serves as counter electrode. Voltages were measured versus an Ag/AgCl reference electrode. In 1~M~\NaSO, pure carbon electrodes were cycled within a voltage window from \mbox{-0.5} to 0.5~V vs. Ag/AgCl, while hybrid electrodes were measured from \mbox{-0.2} to 0.8~V vs. Ag/AgCl.
    Prior to each measurement, the electrodes were vacuum infiltrated and cycled several times at 10~mVs$^{-1}$ until the voltammogram of subsequent cycles showed less than 3\% deviation from the preceding ones.
    Afterwards, at every scan rate 3 cycles were measured in total but only the last one was used for analysis, in order to guarantee stable conditions during the measurement. 
    The gravimetric differential capacitance $C$ was calculated from the measured current $I$ and the applied scan rate $\mrm{d}U/\mrm{d}t$, divided by the total mass of the electrode $m_\mathrm{el}$:
      \begin{equation}
      C = \frac{I}{\mrm{d}U/\mrm{d}t ~ m_\mathrm{el}}.
      \end{equation}

    The total gravimetric capacitance \Cgrav of the electrode is given by the arithmetic mean of the integrated current of the half cycles divided by the measured voltage window. The electrochemical measurements were conducted using an Ivium-n-Stat potentiostat/galvanostat and frequency response analyzer.

\section{Results and Discussion}

  \subsection{Fiber reinforced carbon xerogels}
    Figure \ref{fig:SEM} shows representative SEM micrographs of the samples used in our study. The first row displays the well interconnected network of the bare carbon xerogels with small (S), medium (M) as well as large (L) particle sizes (left to right). Distinct inter-particular meso- and macropores can be observed. 
    
\begin{figure}[ht]
      \begin{minipage}[b]{6cm}
      \subfigure[]{
	\includegraphics[width=9.5cm]{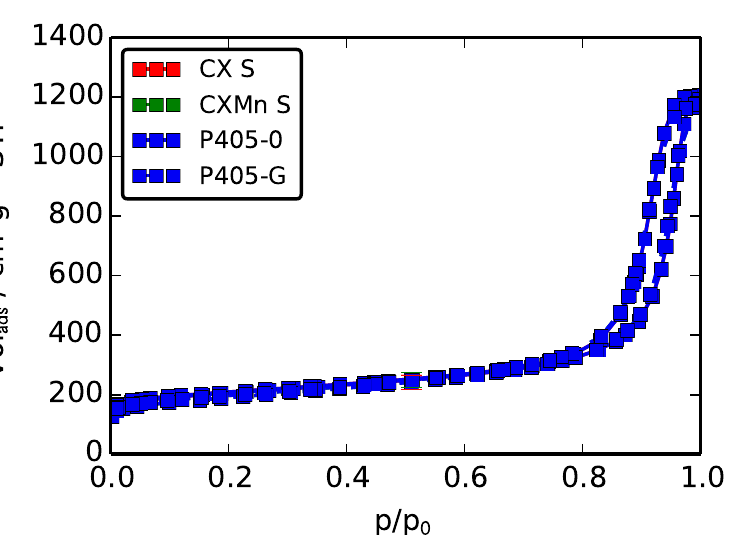}\label{fig:isotherm2}
      }
      \end{minipage}
      \\
      \begin{minipage}[b]{6cm}
      \subfigure[]{
      \includegraphics[width=9.5cm]{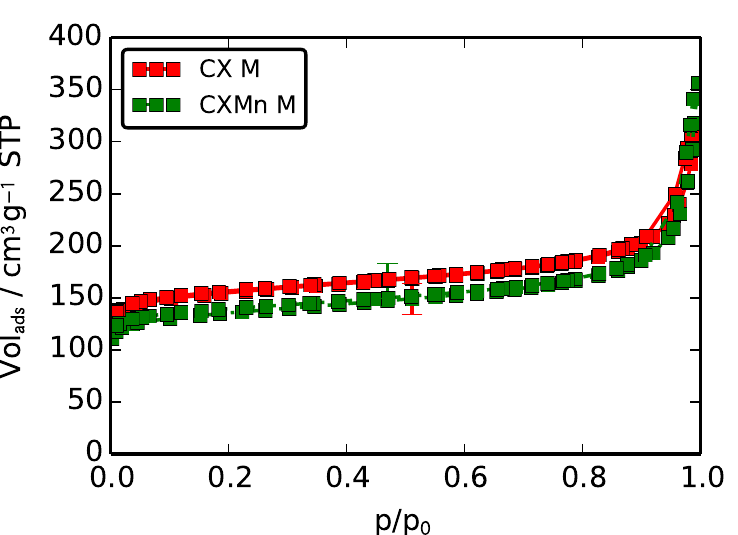}\label{fig:isotherm4}
      }
      \end{minipage}
      \\
      \begin{minipage}[b]{6cm}
      \subfigure[]{
      \includegraphics[width=9.5cm]{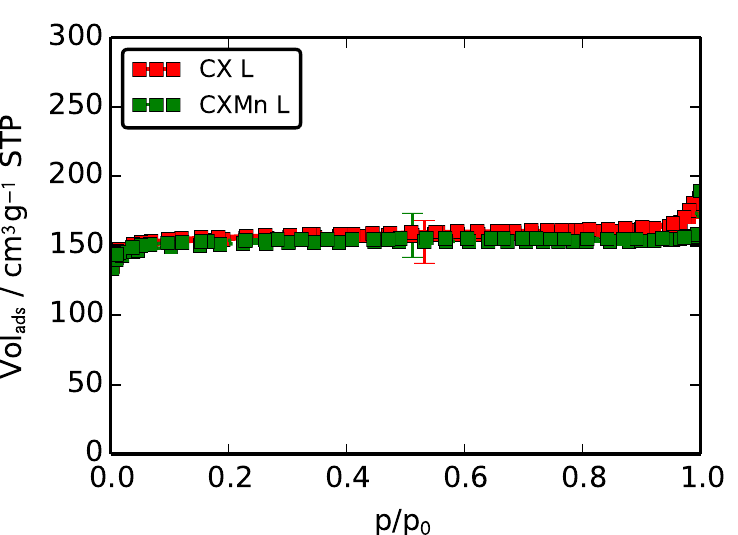}\label{fig:isotherm6}
      }
      \end{minipage}
      \caption{\N-gas sorption isotherms of carbon backbone (red) and hybrid samples (green) for three different particle sizes small (S), medium (M) and large (L), in Figures (a), (b) and (c), respectively. The data are normalized to the carbon xerogel mass without fibers and to standard temperature and pressure (STP).}
      \label{fig:isotherm}
  \end{figure}

    The \N-gas sorption isotherms are shown in Figure \ref{fig:isotherm}. The data are normalized to the mass of the carbon xerogel content of the fiber reinforced material. The isotherms show the typical shape for microporous materials, characterized by a sharp increase in adsorbed volume at relative pressures $p/p_0 < 0.1$, corresponding to micropore filling. In the intermediate relative pressure range, the absorbed volume is larger for samples with small particles than for samples with large particles, indicating an increase in external surface area with decreasing particle size, as expected. For relative pressures above 0.8, a hysteresis loop is observed, which can be related to condensation of \N in mesopores. The hysteresis loop is more pronounced for small particles and decreases for larger particles, hinting at a shift in inter-particular pore size from meso- to macropores and thus to a regime that is not detectable by \N sorption. 
    
    Although BET theory is known to give defective absolute surface area values for microporous carbons, it delivers comparable qualitative results for our samples, assuming the size of micropores being similar for all carbon xerogels under investigation \cite{ IUPAC1984a, Centeno2011, Centeno2010}. This assumption can be rationalized by the identical conditions of sample preparation with respect to the precursor system resorcinol-formaldehyde, curing and pyrolysis step. Additionally, we investigated the microporosity of the carbon xerogels in a separate study, by means of \CO-sorption density functional theory-analysis in combination with small angle X-ray scattering, finding similar micropore size distributions for all resorcinol-formaldehyde derived samples, independently of the primary particle size \cite{Lorrmann2011}.
    
    The results of BET- and t-plot analysis are compiled in Table \ref{tab:data2}. The errors for \Sext for small particle size samples are mainly influenced by the error of about 10~\% in determining the carbon fiber content of the sample. For large particle size samples, the deviation is given by the error of the sorption measurement, with high uncertainties for big interparticular pores. For \SBET and \Vmic, the uncertainty is mainly influenced by the determination of the fiber content, and therefore amounts to 10~\% for all samples. Quantitatively, the external surface area \Sext decreases from 302 m$^2$g$^{-1}$ over 74 m$^2$g$^{-1}$ to 7 m$^2$g$^{-1}$ for the small, medium and large particles, respectively. This trend will be important for interpretation of the results on \MnO-deposition described later. From the gas sorption data, primary particle sizes can be calculated (Eqn. \ref{eqn:dpart}) yielding 12, 53 and 462~nm for samples with small, medium and large particles, respectively \cite{Wiener2004}.
    
\begin{figure}[ht]
  \includegraphics[width=9.5cm]{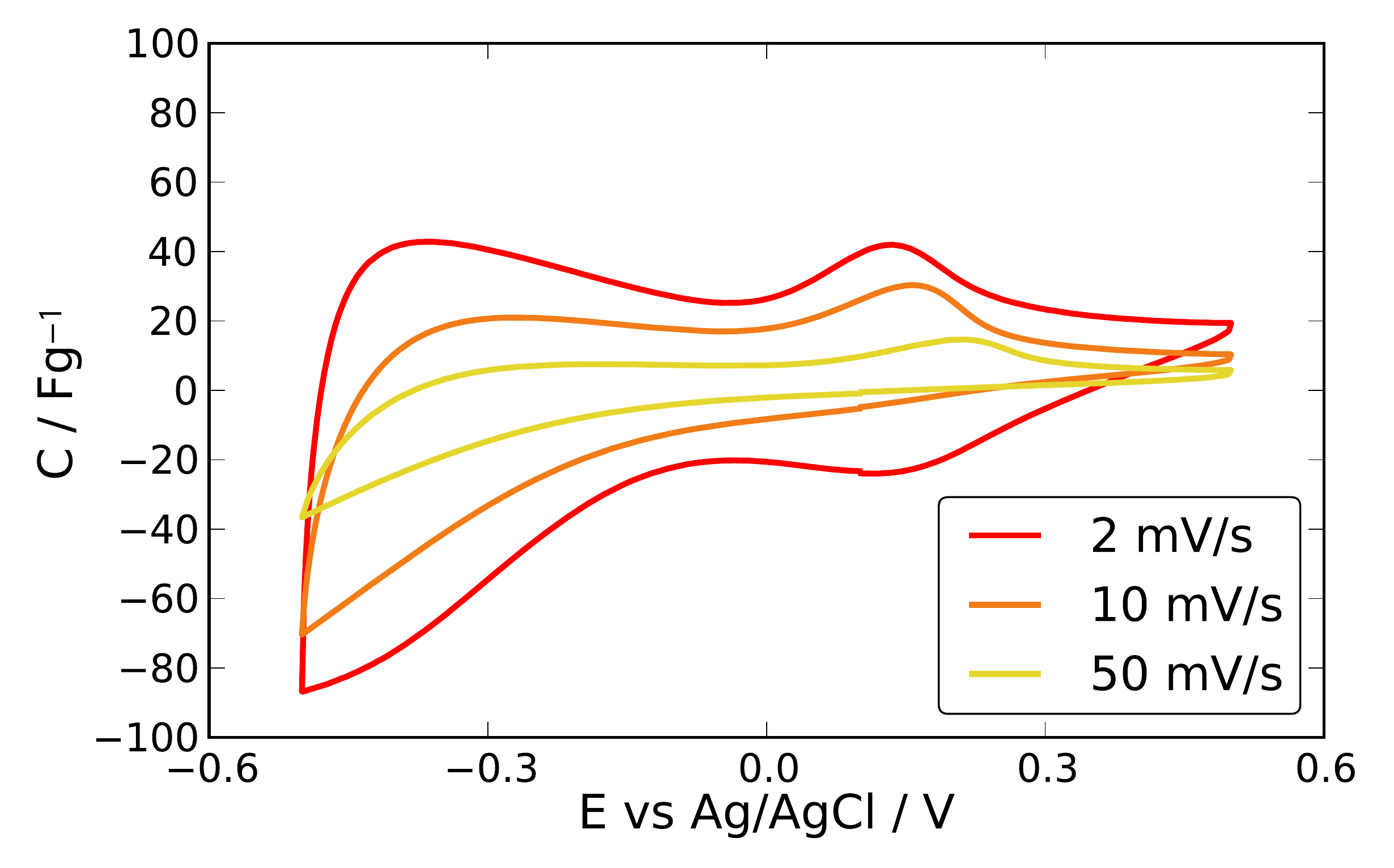}
  \caption{\textbf{Supporting Information:} CV at different scan rates for sample CX~L. } \label{fig:CV_SR_CXL}
  \end{figure}

    Figure \ref{fig:CV_leer} shows the cyclic voltammogram of the fiber reinforced carbon xerogels electrodes.  A butterfly-shaped CV curve with a pronounced  minimum at 0.02~V vs. Ag/AgCl, which is well-known for microporous carbons is observed \cite{Salitra2000, Tobias1983}. With increasing particle size, a sieving effect evolves, indicated by a decrease in the anodic current \cite{Lorrmann2011, Salitra2000, Eliad2001}. Additionally, a bump, becoming more pronounced with increasing particle size, arises. This bump is generally explained by redox reactions of surface groups. However, surface sensitive x-ray photoelectron spectroscopy (XPS) analysis indicates similar groups with respect to their valency for all our resorcinol-formaldehyde derived carbon xerogels.
    We therefore assume that during charging, \SO-ions are forced into pores until all accessible storage sites are occupied, resulting in a broad peak around 0.15~V vs. Ag/AgCl, at 2~mVs$^{-1}$ (see Figure \ref{fig:CV_SR_CXL}, Supporting Information) \cite{Mysyk2009}. These trapped ions are not completely released at the reversed voltage of same absolute value as for trapping; rather the ions become released at higher negative voltages, leading to an increase of negative currents around -0.4~V vs. Ag/AgCl. Another indication for this trapping process is the disappearance of the bump upon ion release with increasing scan rate (Figure \ref{fig:CV_SR_CXL}, Supporting Information).

  \subsection{Fiber reinforced carbon-\MnO xerogels}
    \subsubsection{Variation of deposition conditions}
      
\begin{figure}[ht]
      \begin{minipage}[b]{6cm}
      \subfigure[]{
	\includegraphics[width=9.5cm]{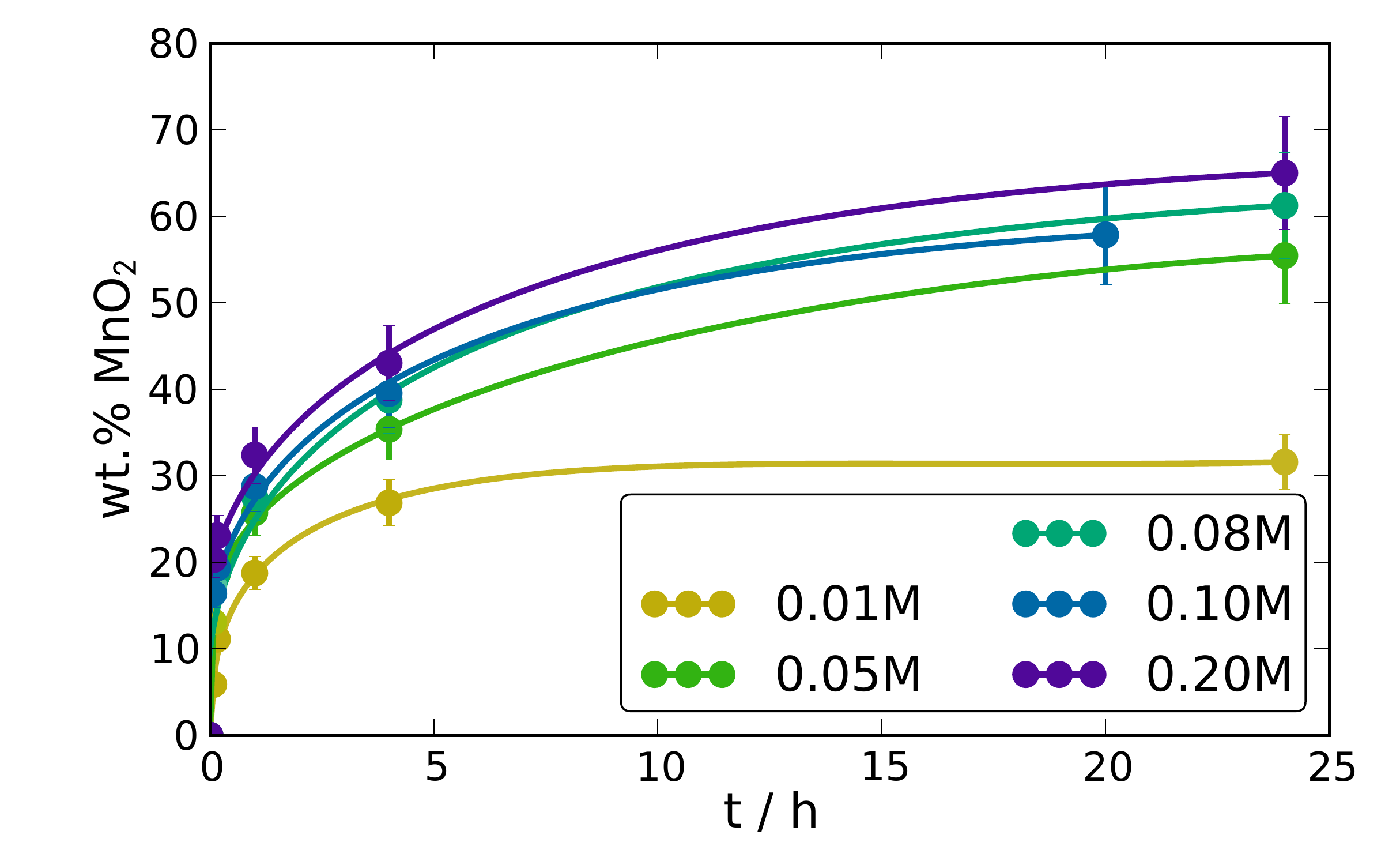}\label{fig:mMnFi}
	}
      \end{minipage}
      \\ 
      \begin{minipage}[b]{6cm}
      \subfigure[]{
	\includegraphics[width=9.5cm]{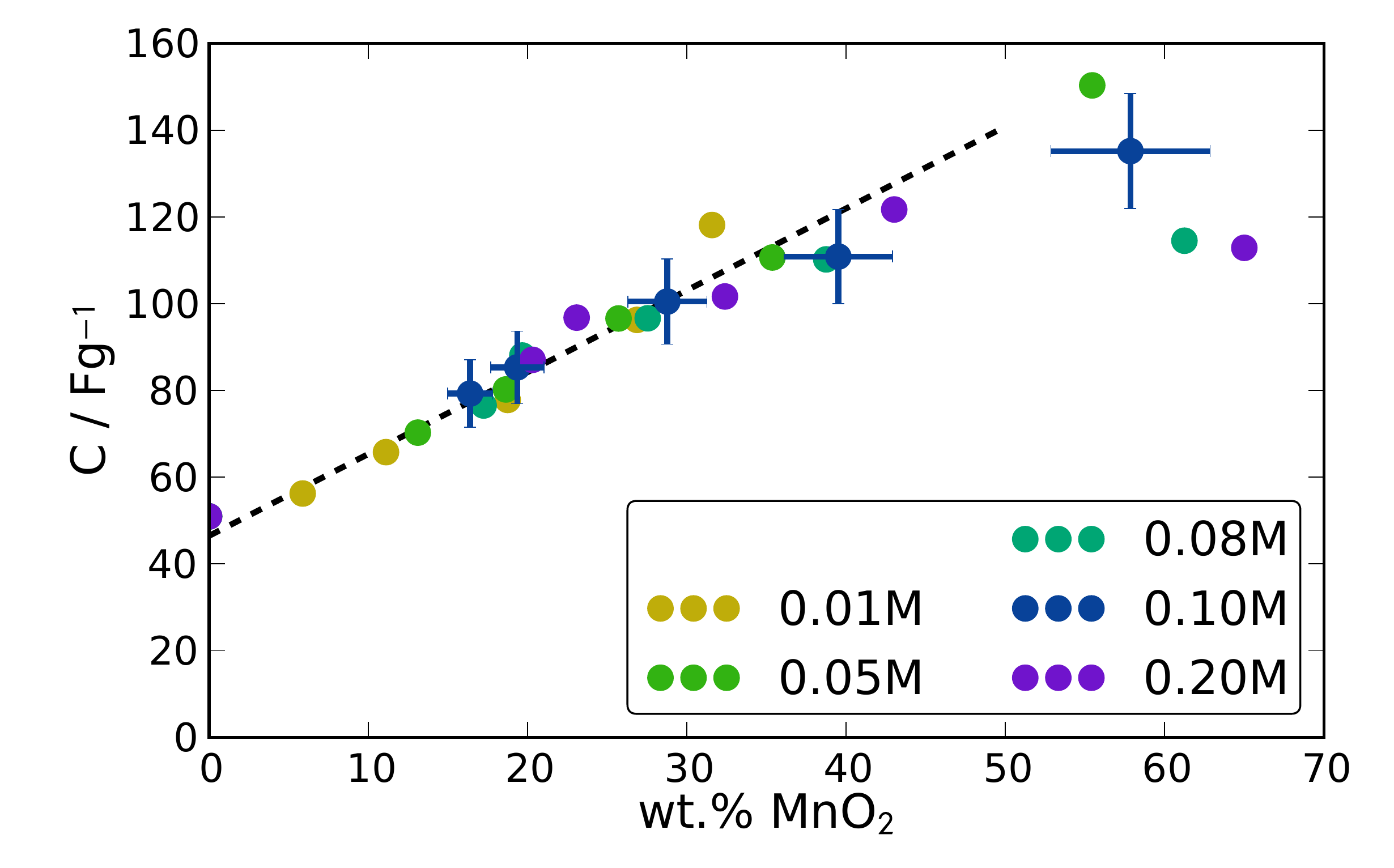}\label{fig:CvsmMnFi}
      }
      \end{minipage}
      \caption{
	(a) Mass percentage of \MnO deposition in the electrodes, plotted against the infiltration time for various concentrations of the \NaMn precursor. Lines are guides to the eye. 
	(b) Specific capacitance for hybrid electrodes as a function of the \MnO uptake. The bare carbon yields a capacitance of about 40 F/g. The dotted line shows a linear fit of the experimental data up to about 35~wt.\% \MnO. Errorbars are drawn exemplarily for the 0.10~M series.}
      \label{fig:mMn}
  \end{figure}

      In a first series of \MnO hybrid electrodes, we varied the concentration of \NaMn precursor and the deposition time, in order to investigate the impact of the synthesis parameters on the resulting electrodes. 
      For convenience and being verified later in this study, from now on manganese oxide refers to \MnO.
      The carbon sample used in these experiments was similar to carbon sample S (Table \ref{tab:data2}), i.e. a fiber reinforced carbon xerogel with a particle size of about 15~nm and external surface area of about 300~m$^2$g$^{-1}$. In Figure \ref{fig:mMnFi} the mass of \MnO per total mass of the electrodes is plotted versus deposition time for various concentrations of the \NaMn precursor. At precursor concentrations between 0.01~M to 0.2~M \NaMn the deposition times were varied from 5 minutes to 24~hours. For the 0.01~M solution, the mass uptake saturates at high deposition times. This can be attributed to a depletion of \Mnion content in the precursor solution, as indicated by the transparency of the solution after about 20~hours. For all samples, a strong increase in mass can be observed at short deposition times. Towards longer deposition times, the slope of the uptake monotonously decreases, but no saturation as expected for a self-limiting process is observed in the applied time range \cite{Fischer2008}. 
      
\begin{figure}[ht]
    \begin{minipage}[b]{6cm}
    \subfigure[]{
      \includegraphics[width=9.5cm]{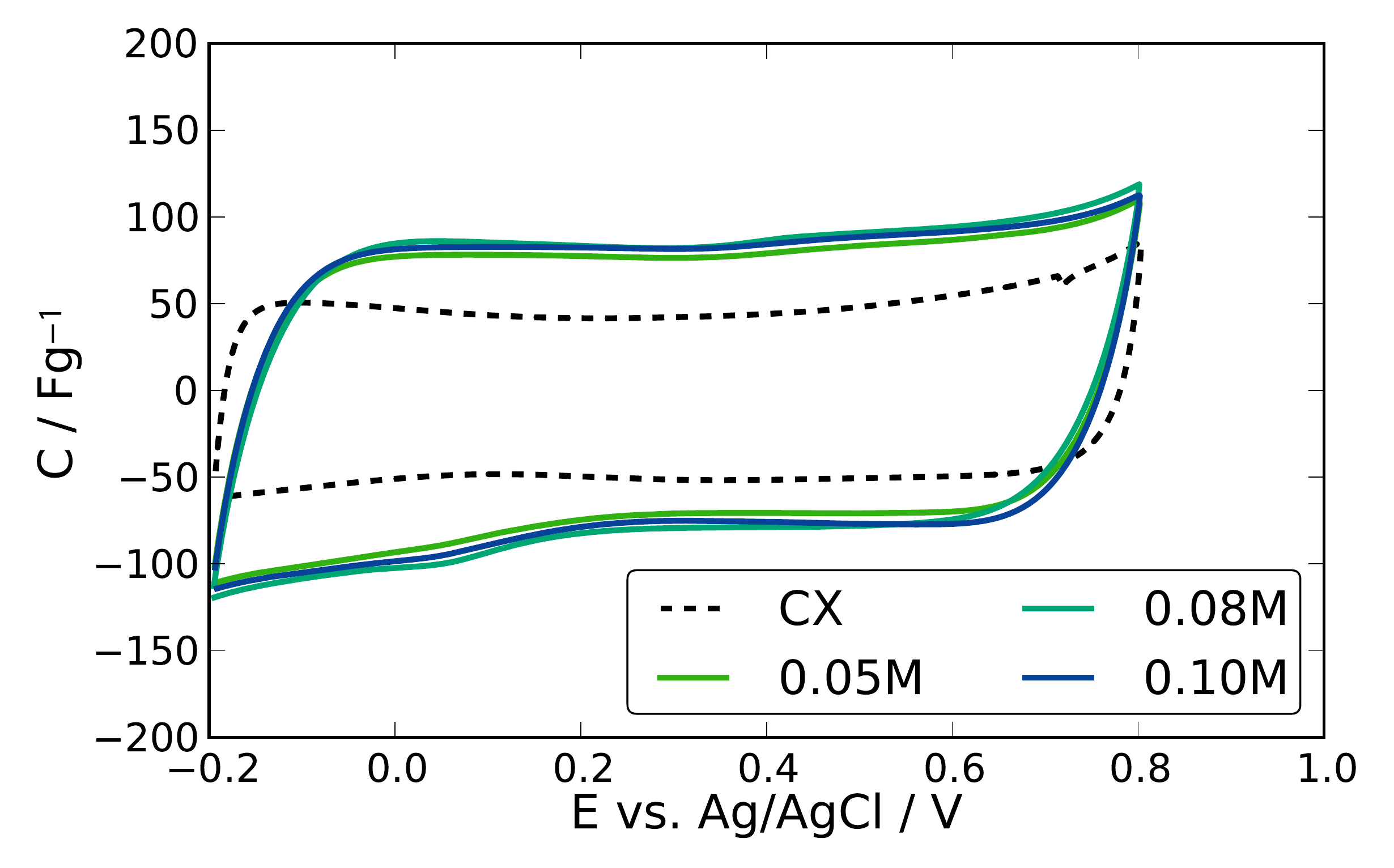}\label{fig:CVFi_A}
      }
      \end{minipage}
      \\ 
      \begin{minipage}[b]{6cm}
      \subfigure[]{
	\includegraphics[width=9.5cm]{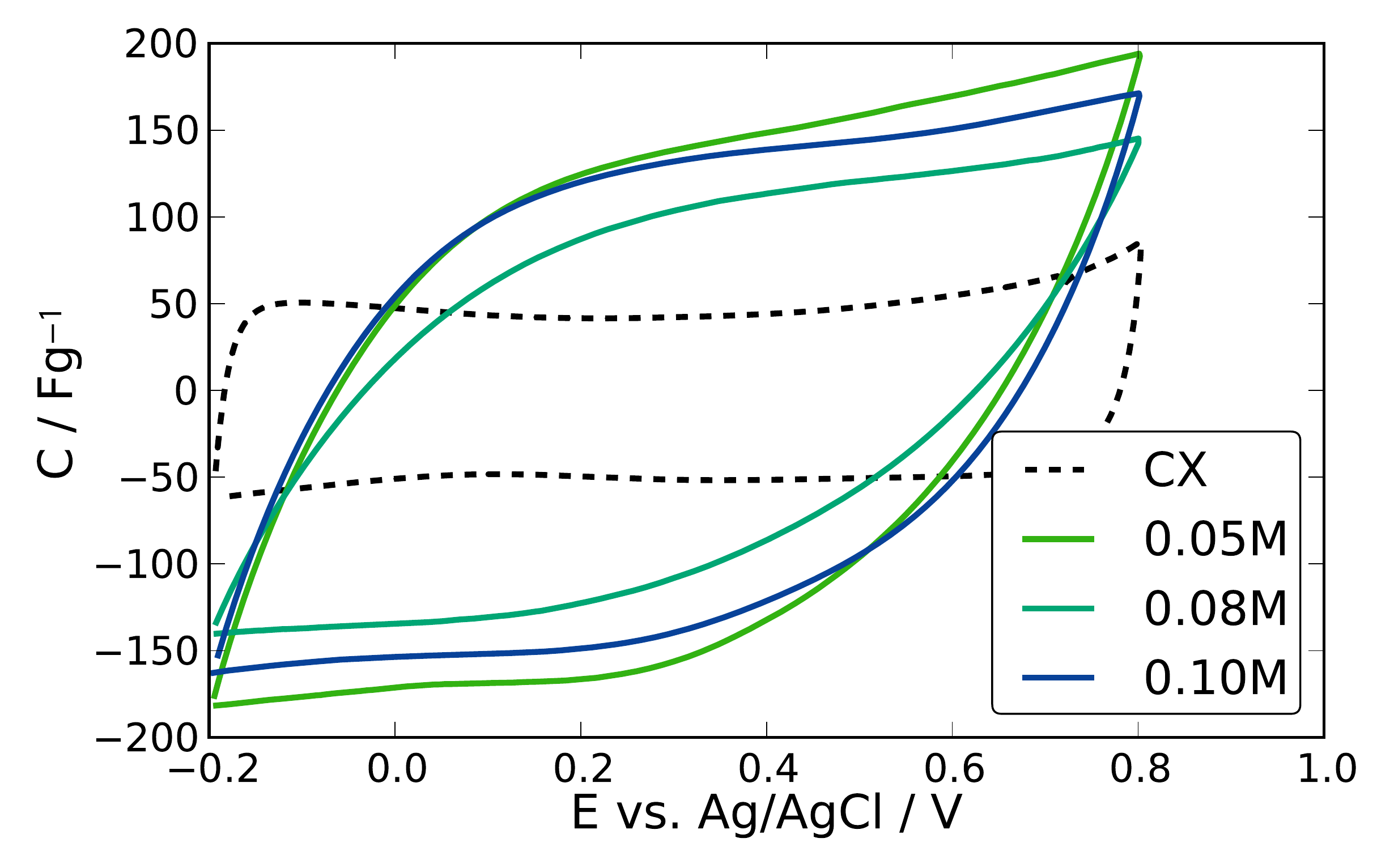}\label{fig:CVFi_B}
	}
	\end{minipage}
	\caption{CV curves for a bare carbon electrode CX and selected samples with a \MnO uptake of about  20~wt.\% (a) and $>$50~wt.\% (b), measured with a potential scan rate of 5 mVs$^{-1}$.}
  \end{figure}

      In Figure \ref{fig:CVFi_A} the CV-curves for samples with an \MnO mass loading of around 20~wt.\% are plotted. Within the range of experimental error, the electrodes show the same electrochemical behavior, thus yielding about the same capacitance value. Samples with a mass loading around 50~wt.\% are plotted in Figure \ref{fig:CVFi_B}. They also contain similar wt.\% of \MnO, but the CV-curve is more disturbed than for samples with a lower \MnO wt.\%. Especially for the 0.08~M electrode, the slope at around -0.2~V vs. Ag/AgCl is less steep, hinting for an increased resistance of the electrode.

      Figure \ref{fig:CvsmMnFi} shows the gravimetric capacitance, calculated from CV measurements at a scan rate of 5~mVs$^{-1}$ for all electrodes shown in Figure \ref{fig:mMnFi}. As one can clearly see, for manganese oxide weight percentages up to about 35~\%, the capacitance only depends on the mass of \MnO that was deposited, rather than on the concentration of \NaMn in the precursor or the deposition time. In other words, the microscopic process of \MnO infiltration into the electrodes is of negligible importance, only the amount of \MnO matters. 

      For mass uptakes above 50~wt.\% however, the precursor concentration influences the capacitance significantly. Whilst for the 0.05~M sample a continuation of the linear trend can be seen, a pronounced deviation occurs for higher \MnO concentrations. These results can be interpreted by a higher amount of \MnO that is not actively participating in charge storage for the samples prepared at precursor concentrations higher than 0.05~M. In this case, the additional \MnO only adds mass to the electrode. From a structural point of view, this \MnO might be deposited as thick layers or agglomerates being inaccessible for the electrolyte. For the 0.05~M sample however, due to the lower concentration of \MnOIon-ions, deposition is assumed to take place more slowly, allowing a more homogeneous distribution of \MnO across the electrode. Further investigations on this microscopic model by small angle x-ray diffraction are forthcoming.
      
      Recently, molecular dynamics simulations have shown that in NaCl aqueous solution ions cannot enter pores smaller than 1.2~nm if no external voltage is applied \cite{Kalluri2011}. Therefore, the assumption proposed by Fischer \textit{et al.} \cite{Fischer2008} that sulfate ions enter micropores by vacuum infiltration and thereafter can be exchanged by \Mnion-ions, leading to deposition of \MnO in the micropores, seems to require further evidence. Vacuum infiltration will only lead to wetting of the micropores by water, but neither \SO nor \Mnion will be able to enter the micropores without an external driving voltage. Therefore, by electroless deposition, as discussed in this study, \MnO will only be deposited on the external surface area of the carbon particles. 
      
    \subsubsection{Variation of carbon particle size}
      
      To investigate the respective adsorption sites, in a second sample series we deposited \MnO on carbon xerogels of different particle sizes. The parameters for deposition were fixed at a concentration of 0.05~M \NaMn and a deposition time of 4~h. These values were taken from the time- and concentration- dependent study, in which a high mass uptake of 35 wt.\% \MnO together with a reasonable electrochemical performance could be achieved (see Figure \ref{fig:mMnFi}, \ref{fig:CvsmMnFi}).
      
\begin{figure}[ht]
      \includegraphics[width=9.5cm]{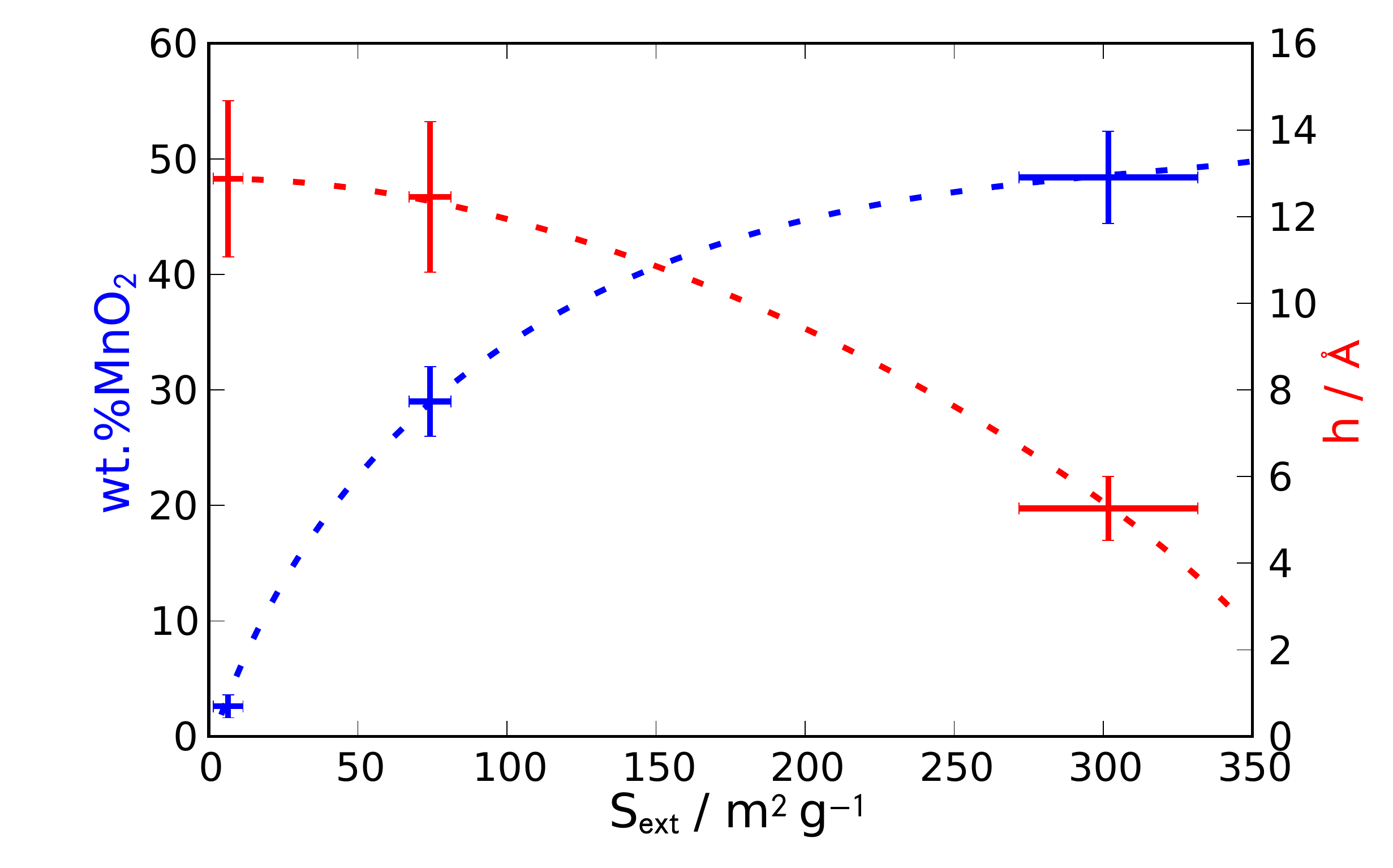}
      \caption{\MnO wt.\% (blue) and \MnO-layer thickness $h$ (red) vs. external surface area. Error bars according to Table \ref{tab:data2} for \Sext and derived from Equation \ref{eqn:h}. The dotted lines serve as a guide to the eye.}\label{fig:mMn_h}
  \end{figure}

      The mass uptake for the S-, M- and L-samples, denoted as CXMn~S, CXMn~M, and CXMn~L, respectively, is plotted as a function of the external surface area of the carbon backbone in Figure \ref{fig:mMn_h} (blue curve). Samples with small particles take up the largest amount of manganese oxide and \textit{vice versa} the dependence on the external surface area \Sext becomes obvious as the sample with the highest external surface area yields the highest mass uptake of \MnO. However, the ratio \mMnOSext is not constant (Figure \ref{fig:mMn_h}), as one would expect in case of a purely self-limiting process. 
      
\begin{figure}[ht]
      \includegraphics[]{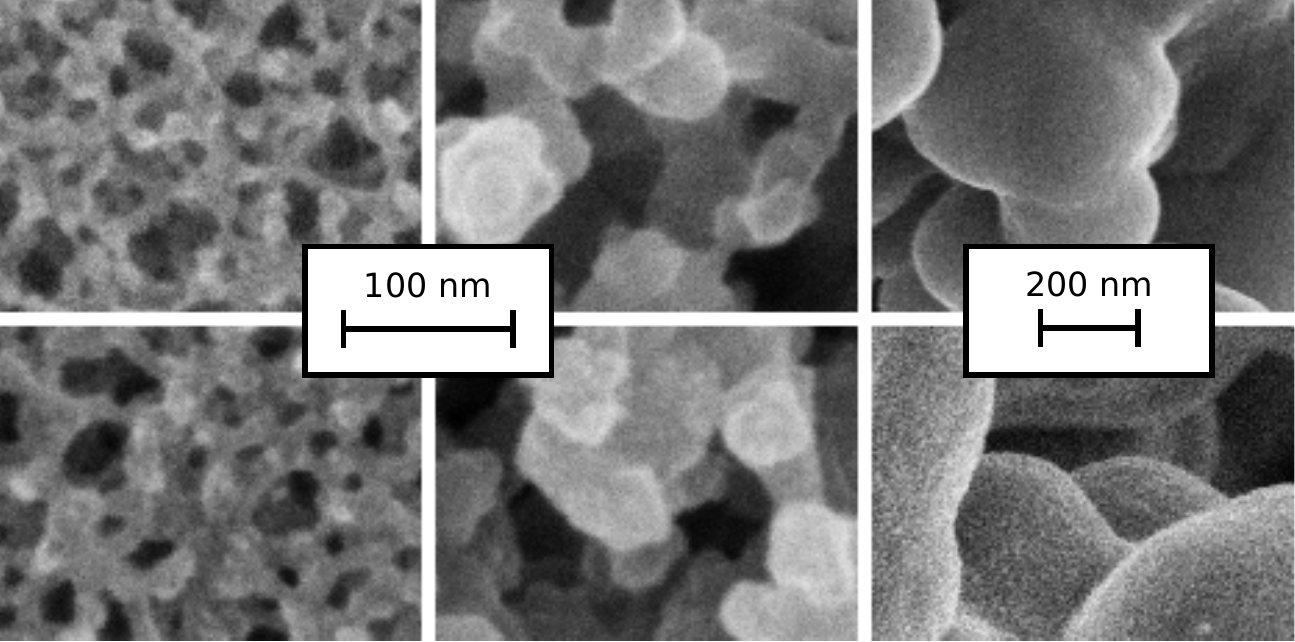}
      \caption{SEM micrographs showing in the upper row the carbon backbone, consisting of small, medium and large particles (left to right). Carbon-Manganese-hybrid samples in the lower row show no significant change in structure. } \label{fig:SEM}
  \end{figure}

      SEM micrographs of the small, medium and large carbon-\MnO samples are shown in the lower row of Figure \ref{fig:SEM}. No significant change in particle size can be detected on the resolved length scales when comparing the hybrid electrode to the corresponding carbon xerogel. Especially the well interconnected network of interparticular pores still remains open and accessible to electrolytic transport. The large particles (right image) seem to be textured on their surface, which may be related to a microstructured \MnO coverage. 
      
\begin{figure}[ht]
  \begin{minipage}[b]{6cm}
  \subfigure[]{
    \includegraphics[width=9.5cm]{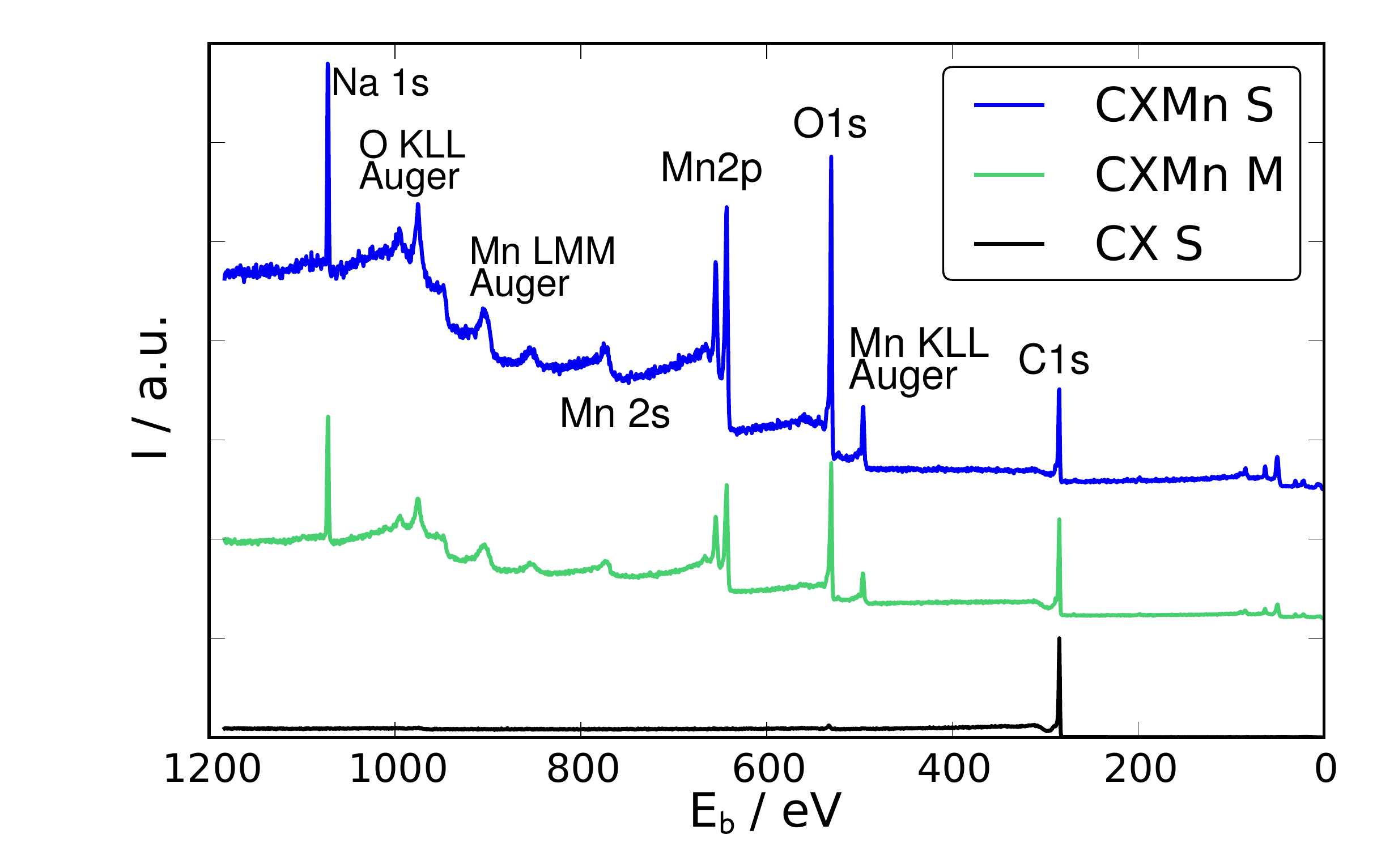}\label{fig:XPS}
    }
    \end{minipage}
    \\ 
    \begin{minipage}[b]{6cm}
    \subfigure[]{
      \includegraphics[width=9.5cm]{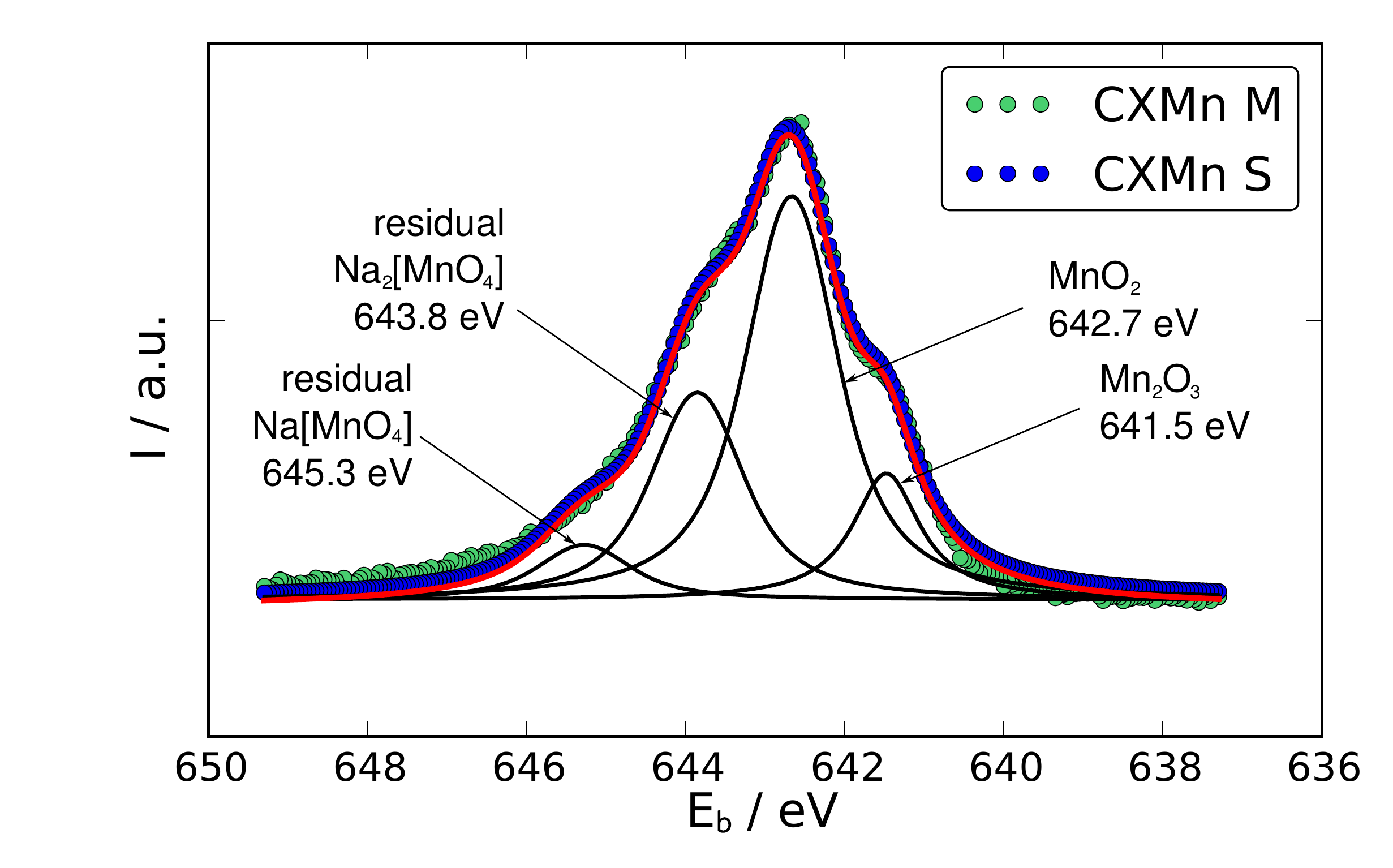}\label{fig:XPS_Mn2p}
      }
      \end{minipage}
      \caption{\textbf{Supporting Information:} Overview of XPS data, intensities normalized to the C1s peak and shifted for clarity: Pure carbon CX~S and hybrid samples CXMn~M, CXMn~S (a). XPS-peak of Mn2p$_{3/2}$ for samples CXMn~S and CXMn~M (b), modeling of peaks according to \cite{Iwanowski2004, Oku1996}. The data for CXMn~M was scaled by a factor of 2.1.}
  \end{figure}

      Via XPS measurements, samples of small and medium particle sizes proved to be chemically identical (see Figure \ref{fig:XPS}, Supporting Information). A fit of the Mn2p$_{3/2}$-peak (see Figure \ref{fig:XPS_Mn2p}, Supporting Information) reveals the manganese oxide in the electrode to be a mixture of Mn$_2$O$_3$ and \MnO at a ratio of 1:4.6 \cite{Jiang2002, Iwanowski2004, Oku1996}. From this analysis we conclude the manganese oxide prepared in this study to consist preferentially of \MnO.
      The XPS analysis also hints for residual \NaSO and Na$_2$SO$_4$ in the samples. However, since all samples were processes by an identical route of preparation and in particular were subjected to the same washing procedure, the ratio of sodium sulfate- to manganese oxide-species can be assumed similar for all samples. 
      
      The gas sorption isotherms normalized to the total mass of the electrodes show a decrease in adsorbed volume which is of the order of the weight percentage of the deposited \MnO, namely 48, 29 and 3~wt.\% for small, medium and large particles, respectively. Therefore, no substantial fraction of pore volume (detectable by \N sorption) was secluded or filled with \MnO, rather the material is contributing to the sample mass only. In order to emphasize this effect more clearly, the isotherms normalized to the CX mass in the sample are plotted in Figure \ref{fig:isotherm} and compared to the respective isotherms of the bare CX.
      For higher relative pressures the drop in the adsorbed volume differs significantly from the mass uptake, thus corresponding to a decrease of meso- and macropore volume in the course of \MnO deposition on the external surface area of the carbon particles. This trend becomes also obvious in the calculated values of external surface area \Sext and the micropore volume \Vmic (see Table \ref{tab:data2}). The small particles show the biggest decrease (40~\%) in \Sext, due to the fact that most of the \MnO was deposited on the external surface area of the carbon particles. As the pronounced decrease in micropore volume (64~\%) can not only be attributed to a gain in mass, a significant number of micropores were either partially filled with \MnO or closed, hence are no longer accessible to \N molecules.	
      
      The calculated values (Table \ref{tab:data2}) from gas sorption data reveal an increase in \Sext for all particle sizes if normalized to the mass of carbon xerogel in the sample.       
      The external surface area shows a slight systematic increase, indicating additional changes of the microstructure induced by the \MnO deposit.
      
      Supposed all mass of \MnO being deposited on the external surface area of the bare carbon xerogels, we can estimate the layer thickness $h$ by using a bulk density of \rhoMnO~=~5.026~gcm$^{-3}$ \cite{DichteMnO2}:
	\begin{equation}
	h = \frac{m_{\mathrm{MnO_2}}}{m_\mathrm{CX}} \cdot \rho_{\mathrm{MnO_2}}^{-1} \cdot S_\mathrm{ext}^{-1}.\label{eqn:h}
	\end{equation}
      The resulting layer thickness is plotted vs. the external surface area in Figure \ref{fig:mMn_h} (red curve). As one can see, the samples with the largest particle size take up the least amount of \MnO, however resulting in the thickest \MnO-layer, of about 13 \AA. This allows for two possible interpretations: The deposited \MnO mass was partially deposited as agglomerates within the carbon structure. SEM overview scans however do not show any irregularities in the carbon structure (see Figure \ref{fig:SEM}).
      Therefore, an additional competitive deposition mechanism besides the previously proposed self-limiting process might lead to thicker deposits on the carbon surface area.
	
\begin{figure}[ht]
  \begin{minipage}[b]{6cm}
  \subfigure[]{
    \includegraphics[width=9.5cm]{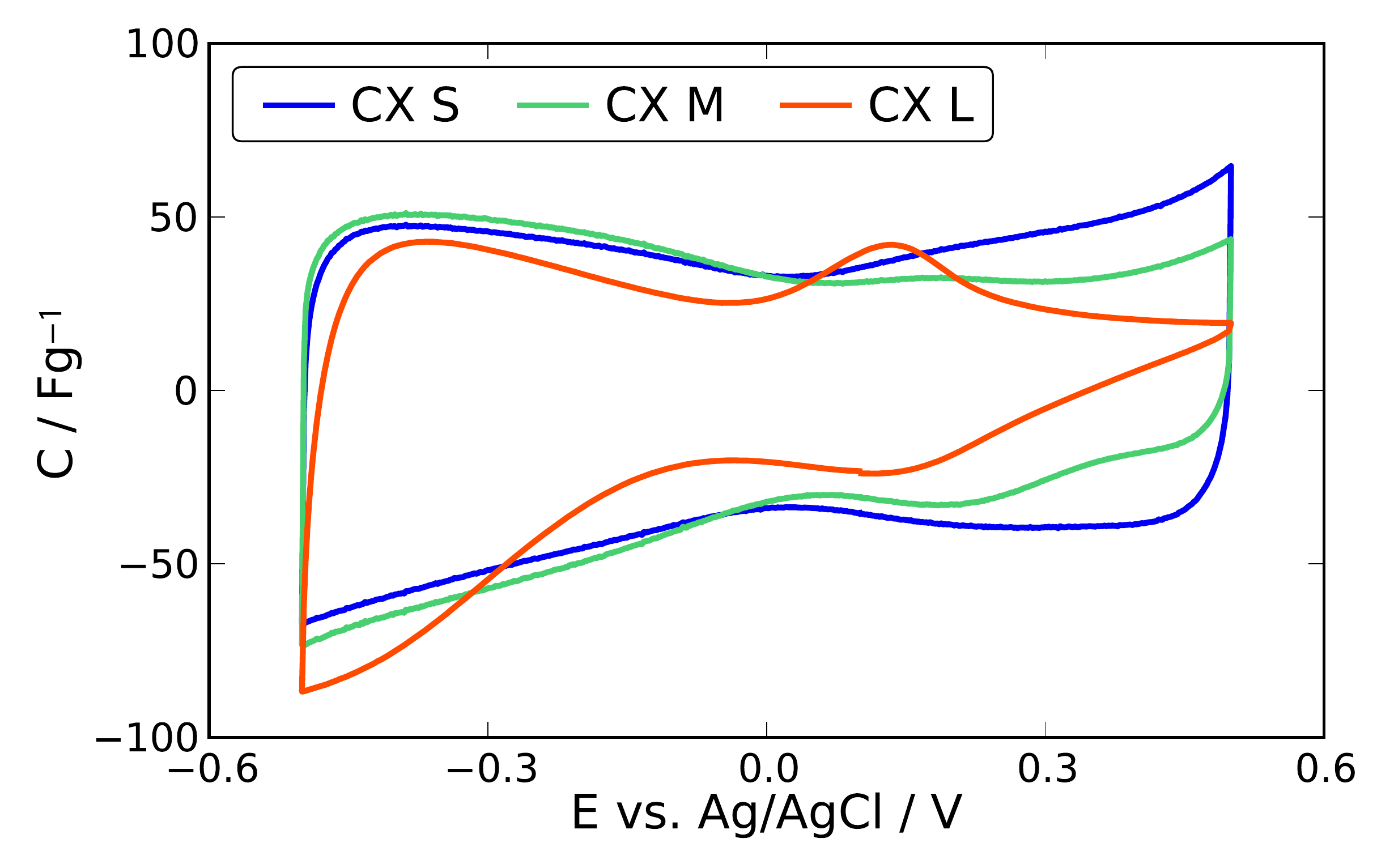}\label{fig:CV_leer}
    }
    \end{minipage}
  \\ 
    \begin{minipage}[b]{6cm}
    \subfigure[]{
      \includegraphics[width=9.5cm]{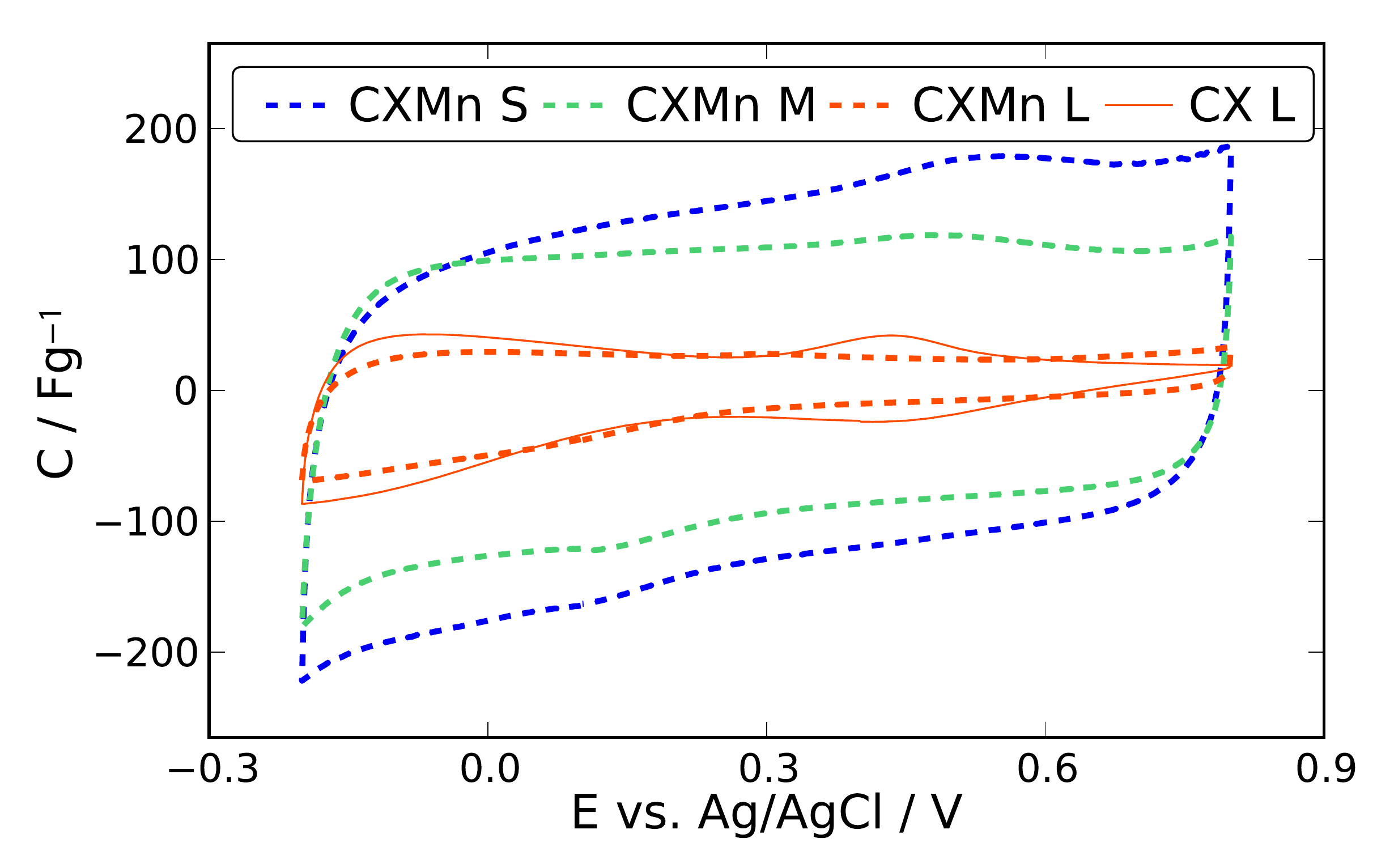}\label{fig:CV_gef}
      }
      \end{minipage}
      \caption{Cyclic voltammogram at 2~mVs$^{-1}$ of bare carbon samples (full lines) with different particles sizes (a) and corresponding hybrid electrodes (dashed lines) (b).}\label{fig:CV}
  \end{figure}

      The cyclic voltammograms for the hybrid electrodes are displayed in Figure \ref{fig:CV_gef}. 
      For large particle sizes, the gain in gravimetric capacitance by \MnO pseudocapacitance is compensated by the gain in mass. In contrast, hybrid electrodes with medium and small particle sizes show an huge increase in capacitance while the corresponding CV curves still retain a rectangular shape, only marginally influenced by the reduced conductivity due to the \MnO layer formation. The bumps at about 0.1~V and 0.5~V~vs.~Ag/AgCl for samples composed of medium and small particles in Figure \ref{fig:CV_gef} are an exclusive feature of the hybrid electrodes and have been attributed to ionic intercalation processes into \MnO upon charging \cite{Kanoh1997}.
	
\begin{figure}[ht]
      \begin{minipage}[b]{6cm}
      \subfigure[]{
      \includegraphics[width=9.5cm]{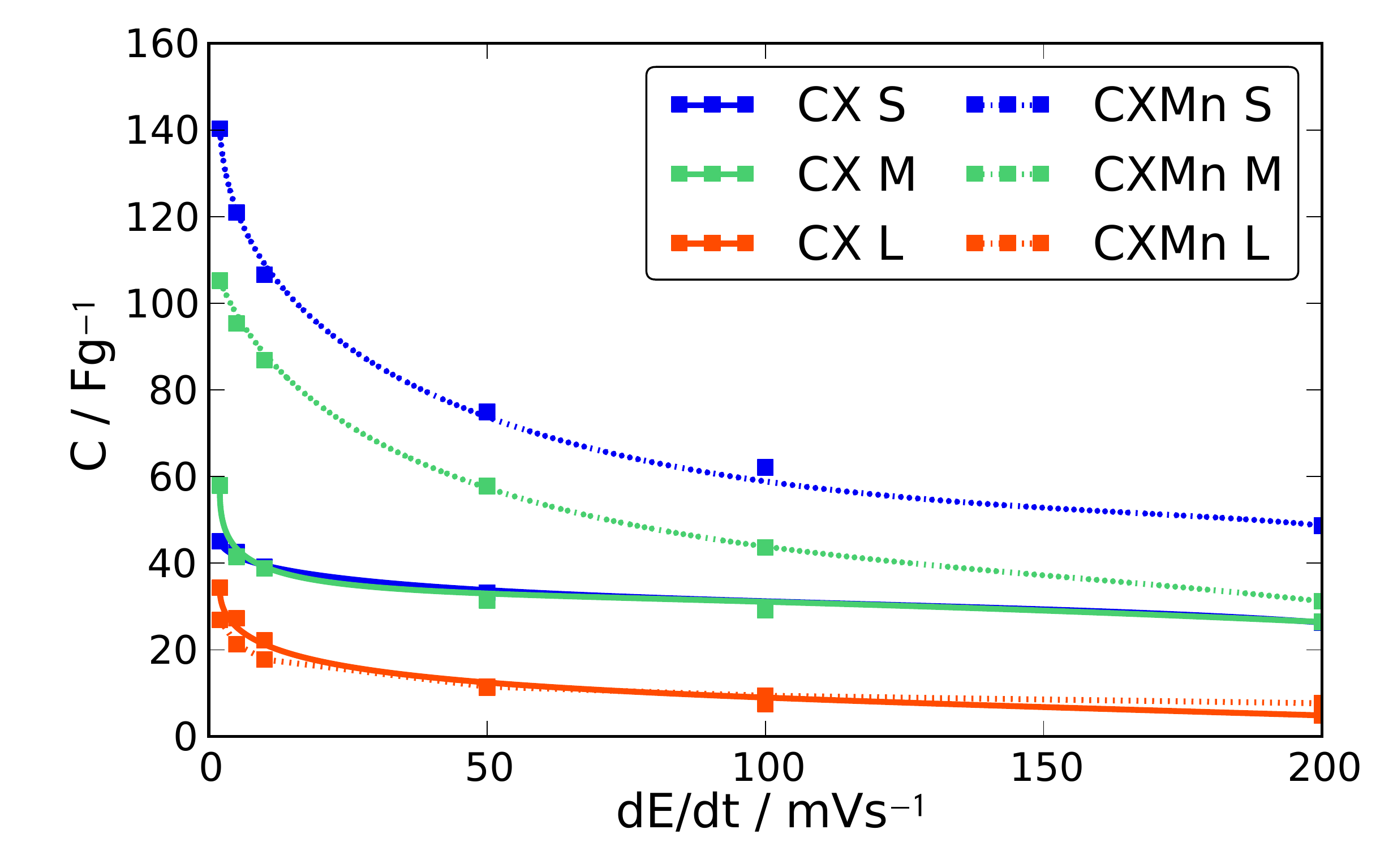}\label{fig:CV_CvsSR_alle}
      }
      \end{minipage}
      \\ 
      \begin{minipage}[b]{6cm}
      \subfigure[]{
      \includegraphics[width=9.5cm]{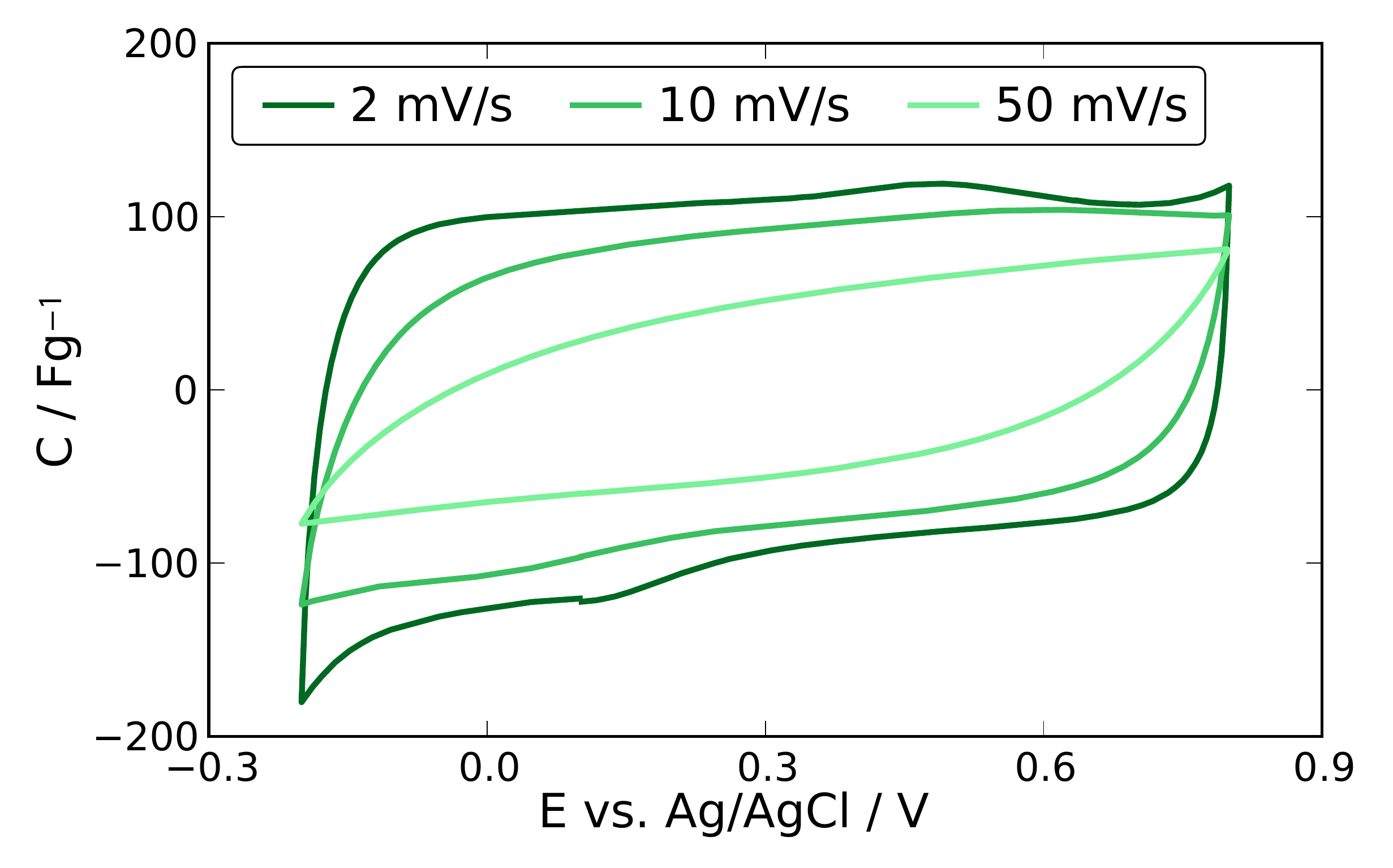}\label{fig:CV_CvsSR_m}
      }
      \end{minipage}
      \caption{(a) Gravimetric capacitance vs. applied scan rate, carbon samples (solid lines) and hybrid samples (dashed lines). (b) Cyclic voltammogram of sample CX~Mn~M for different scan rates.}\label{fig:CV_CvsSR}
  \end{figure}

      In Figure \ref{fig:CV_CvsSR_alle} the gravimetric capacitance is plotted against the applied scan rate. Within the measured range, the capacitance of the carbon samples is less dependent on the scan rate compared to the hybrid electrodes. At 100~mVs$^{-1}$ the capacitance of the latter drops by 56~\% for CXMn~S and by 33~\% for CXMn~M. The reason for this behavior can be elucidated by comparing the CV-curves at the different scan rates, for instance for sample CXMn~M, plotted in Figure \ref{fig:CV_CvsSR_m}. The rectangular shape at 2~mVs$^{-1}$ is increasingly disturbed at higher scan rates, mainly due to the decreasing slope at -0.3~V~vs.~Ag/AgCl. These observations can be attributed to the strong influence of the low conductive \MnO layer, both at the carbon-\MnO-electrolyte interface and at the contact between the current collector and the carbon backbone.
      
      Although the samples made of small particles the hybrid electrodes show a strong dependence of the capacitance on the scan rate, at 200~mVs$^{-1}$ the capacitance is still about 2 times higher than that of the bare carbon electrode.

\section{Conclusion}
	
  Our study of the carbon-\MnO hybrid model system addresses the incorporation of \MnO into resorcinol-formaldehyde-derived carbon xerogels exhibiting different primary particle sizes and therefore different external surface area. A series of various infiltration times and concentrations showed no dependence on the preparation parameters for mass uptakes of \MnO up to 35~wt.\%. For high mass uptakes, low concentrations of \NaMn and long deposition times result in the largest capacitance increase by a factor of 3. By varying the size of the primary carbon particles in the xerogel, we demonstrated that deposition mainly takes place on the external surface area of the carbon backbone. Small backbone particles provide a higher specific external surface area and therefore can host more \MnO per mass of carbon xerogel than large particles. 
  
  \subsection*{Acknowledgements}

    Funding by Deutsche Bundesstiftung Umwelt is gratefully acknowledged. The authors thank Matthias Wiener (ZAE Bayern) for support with \N~sorption measurements and Andreas Ruff, Sebastian Meyer and Ralph Claessen (Exp. Physics 4, University of Wuerzburg) for assistance on XPS measurements.

\newpage
\cleardoublepage

\section{Tables}
    \begin{table}[h]
    \caption{Overview of structural data derived for the carbon xerogels as well as the corresponding hybrid xerogels. Mass specific values are normalized to the mass of carbon xerogel, the density $\rho$ includes fibers and structural inhomogeneities in the electrodes. }\label{tab:data2}
    \begin{center}
    \begin{tabular}{l|cc|ccc|cc}
         & wt.\% & $\rho$  & \Sext  & S$_\mrm{{BET}}$  & \Vmic    &  \dpart     & \dpore \\
    sample & \MnO & [gcm$^{-3}$] & [m$^2$g$^{-1}$] & [m$^2$g$^{-1}$] &  [cm$^3$g$^{-1}$] &   [nm] & [nm]\\
    \hline
    CX S & 0     & 0.24$\pm$0.02 & 302$\pm$30 & 751$\pm$75 & 0.18$\pm$0.02 & 12$\pm$1    & 14$\pm$4  \\
    CX M & 0     & 0.16$\pm$0.02 & 74 $\pm$7  & 601$\pm$60 & 0.21$\pm$0.02 & 53$\pm$6    & 108$\pm$31 \\
    CX L & 0     & 0.15$\pm$0.02 & 7  $\pm$5  & 623$\pm$62 & 0.24$\pm$0.02 & 462$\pm$49  & 1184$\pm$355\\
    \hline
    CXMn S & 48$\pm$4  & 0.39$\pm$0.04 & 350$\pm$35 & 685$\pm$69 & 0.15$\pm$0.02 & --  & --  \\
    CXMn M & 29$\pm$3  & 0.25$\pm$0.03 & 78 $\pm$8  & 528$\pm$53 & 0.18$\pm$0.02 & --  & -- \\
    CXMn L & 3$\pm$1   & 0.21$\pm$0.02 & 1  $\pm$5  & 610$\pm$61 & 0.24$\pm$0.02 & --  & --
    \end{tabular}
    \end{center}
    \end{table}
    
    \newpage
    \clearpage
    \cleardoublepage

\section{Figures}

  \newpage
  \clearpage

\newpage
\clearpage
\cleardoublepage

\section*{References}
\bibliography{paper_carbon} 

\begin{thebibliography}{36}
\providecommand{\natexlab}[1]{#1}
\providecommand{\url}[1]{\texttt{#1}}
\providecommand{\urlprefix}{URL }
\expandafter\ifx\csname urlstyle\endcsname\relax
  \providecommand{\doi}[1]{doi:\discretionary{}{}{}#1}\else
  \providecommand{\doi}{doi:\discretionary{}{}{}\begingroup
  \urlstyle{rm}\Url}\fi
\providecommand{\eprint}[2][]{\url{#2}}
\providecommand{\BIBand}{and}
\providecommand{\bibinfo}[2]{#2}
\ifx\xfnm\undefined \def\xfnm[#1]{\unskip,\space#1}\fi
\bibitem[{Conway(1999)}]{Conway1999}
\bibinfo{author}{Conway\xfnm[ B.E.]}.
\newblock \bibinfo{title}{Electrochemical Supercapacitors: Scientific
  Fundamentals and Technological Applications}.
\newblock \bibinfo{address}{New York}: \bibinfo{publisher}{Kluwer Academic /
  Plenum Publishers}; \bibinfo{edition}{1.} ed.; \bibinfo{year}{1999}.
\bibitem[{Miller and Burke(2008)}]{Miller2008}
\bibinfo{author}{Miller\xfnm[ J.R.]}, \bibinfo{author}{Burke\xfnm[ A.F.]}.
\newblock \bibinfo{title}{{Electrochemical capacitors: Challenges and
  opportunities for real-world applications}}.
\newblock \bibinfo{journal}{Electrochemical Society Interface}
  \bibinfo{year}{2008};\bibinfo{volume}{17}(\bibinfo{number}{1}):\bibinfo{pages}{53--57}.
\bibitem[{K\"{o}tz and Carlen(2000)}]{Koetz2000}
\bibinfo{author}{K\"{o}tz\xfnm[ R.]}, \bibinfo{author}{Carlen\xfnm[ M.]}.
\newblock \bibinfo{title}{{Principles and applications of electrochemical
  capacitors}}.
\newblock \bibinfo{journal}{Electrochimica Acta}
  \bibinfo{year}{2000};\bibinfo{volume}{45}(\bibinfo{number}{15-16}):\bibinfo{pages}{2483--2498}.
\bibitem[{Frackowiak and Beguin(2001)}]{Frackowiak2001}
\bibinfo{author}{Frackowiak\xfnm[ E.]}, \bibinfo{author}{Beguin\xfnm[ F.]}.
\newblock \bibinfo{title}{{Carbon materials for the electrochemical storage of
  energy in capacitors}}.
\newblock \bibinfo{journal}{Carbon}
  \bibinfo{year}{2001};\bibinfo{volume}{39}(\bibinfo{number}{6}):\bibinfo{pages}{937--950}.
\bibitem[{Lee and Goodenough(1999)}]{Lee1999}
\bibinfo{author}{Lee\xfnm[ H.Y.]}, \bibinfo{author}{Goodenough\xfnm[ J.B.]}.
\newblock \bibinfo{title}{{Supercapacitor Behavior with KCl Electrolyte}}.
\newblock \bibinfo{journal}{Journal of Solid State Chemistry}
  \bibinfo{year}{1999};\bibinfo{volume}{144}(\bibinfo{number}{1}):\bibinfo{pages}{220--223}.
\bibitem[{Toupin et~al.(2004)Toupin, Brousse and B\'{e}langer}]{Toupin2004}
\bibinfo{author}{Toupin\xfnm[ M.]}, \bibinfo{author}{Brousse\xfnm[ T.]},
  \bibinfo{author}{B\'{e}langer\xfnm[ D.]}.
\newblock \bibinfo{title}{{Charge Storage Mechanism of MnO 2 Electrode Used in
  Aqueous Electrochemical Capacitor}}.
\newblock \bibinfo{journal}{Chemistry of Materials}
  \bibinfo{year}{2004};\bibinfo{volume}{16}(\bibinfo{number}{16}):\bibinfo{pages}{3184--3190}.
\bibitem[{Beaudrouet et~al.(2009)Beaudrouet, Legallasalle and
  Guyomard}]{Beaudrouet2009}
\bibinfo{author}{Beaudrouet\xfnm[ E.]}, \bibinfo{author}{Legallasalle\xfnm[
  A.]}, \bibinfo{author}{Guyomard\xfnm[ D.]}.
\newblock \bibinfo{title}{{Nanostructured manganese dioxides: Synthesis and
  properties as supercapacitor electrode materials}}.
\newblock \bibinfo{journal}{Electrochimica Acta}
  \bibinfo{year}{2009};\bibinfo{volume}{54}(\bibinfo{number}{4}):\bibinfo{pages}{1240--1248}.
\bibitem[{Jacob et~al.(2009)Jacob, Yang and Zhitomirsky}]{Jacob2009}
\bibinfo{author}{Jacob\xfnm[ G.M.]}, \bibinfo{author}{Yang\xfnm[ Q.M.]},
  \bibinfo{author}{Zhitomirsky\xfnm[ I.]}.
\newblock \bibinfo{title}{{Composite electrodes for electrochemical
  supercapacitors}}.
\newblock \bibinfo{journal}{Journal of Applied Electrochemistry}
  \bibinfo{year}{2009};\bibinfo{volume}{39}(\bibinfo{number}{12}):\bibinfo{pages}{2579--2585}.
\bibitem[{Reddy and Reddy(2004)}]{Reddy2004}
\bibinfo{author}{Reddy\xfnm[ R.N.]}, \bibinfo{author}{Reddy\xfnm[ R.G.]}.
\newblock \bibinfo{title}{{Synthesis and electrochemical characterization of
  amorphous MnO2 electrochemical capacitor electrode material}}.
\newblock \bibinfo{journal}{Journal of Power Sources}
  \bibinfo{year}{2004};\bibinfo{volume}{132}(\bibinfo{number}{1-2}):\bibinfo{pages}{315--320}.
\bibitem[{Cross et~al.(2011)Cross, Morel, Hollenkamp and Donne}]{Cross2011}
\bibinfo{author}{Cross\xfnm[ A.D.]}, \bibinfo{author}{Morel\xfnm[ A.]},
  \bibinfo{author}{Hollenkamp\xfnm[ T.F.]}, \bibinfo{author}{Donne\xfnm[
  S.W.]}.
\newblock \bibinfo{title}{{Chronoamperometric Versus Galvanostatic Preparation
  of Manganese Oxides for Electrochemical Capacitors}}.
\newblock \bibinfo{journal}{Journal of The Electrochemical Society}
  \bibinfo{year}{2011};\bibinfo{volume}{158}(\bibinfo{number}{10}):\bibinfo{pages}{A1160--A1165}.
\bibitem[{Wei et~al.(2007)Wei, Nagarajan and Zhitomirsky}]{wei2007}
\bibinfo{author}{Wei\xfnm[ J.]}, \bibinfo{author}{Nagarajan\xfnm[ N.]},
  \bibinfo{author}{Zhitomirsky\xfnm[ I.]}.
\newblock \bibinfo{title}{{Manganese oxide films for electrochemical
  supercapacitors}}.
\newblock \bibinfo{journal}{Journal of Materials Processing Technology}
  \bibinfo{year}{2007};\bibinfo{volume}{186}(\bibinfo{number}{1-3}):\bibinfo{pages}{356--361}.
\bibitem[{Dong et~al.(2006)Dong, Shen, Gu, Xiong, Zhu, Li et~al.}]{Dong2006}
\bibinfo{author}{Dong\xfnm[ X.]}, \bibinfo{author}{Shen\xfnm[ W.]},
  \bibinfo{author}{Gu\xfnm[ J.]}, \bibinfo{author}{Xiong\xfnm[ L.]},
  \bibinfo{author}{Zhu\xfnm[ Y.]}, \bibinfo{author}{Li\xfnm[ H.]}, et~al.
\newblock \bibinfo{title}{{A structure of MnO2 embedded in CMK-3 framework
  developed by a redox method}}.
\newblock \bibinfo{journal}{Microporous and Mesoporous Materials}
  \bibinfo{year}{2006};\bibinfo{volume}{91}(\bibinfo{number}{1-3}):\bibinfo{pages}{120--127}.
\bibitem[{Lei et~al.(2008)Lei, Fournier, Pascal and Favier}]{Lei2008}
\bibinfo{author}{Lei\xfnm[ Y.]}, \bibinfo{author}{Fournier\xfnm[ C.]},
  \bibinfo{author}{Pascal\xfnm[ J.L.]}, \bibinfo{author}{Favier\xfnm[ F.]}.
\newblock \bibinfo{title}{{Mesoporous carbon manganese oxide composite as
  negative electrode material for supercapacitors}}.
\newblock \bibinfo{journal}{Microporous and Mesoporous Materials}
  \bibinfo{year}{2008};\bibinfo{volume}{110}(\bibinfo{number}{1}):\bibinfo{pages}{167--176}.
\bibitem[{Lin et~al.(2011)Lin, Wei, Chien and Lu}]{Lin2011}
\bibinfo{author}{Lin\xfnm[ Y.H.]}, \bibinfo{author}{Wei\xfnm[ T.Y.]},
  \bibinfo{author}{Chien\xfnm[ H.C.]}, \bibinfo{author}{Lu\xfnm[ S.Y.]}.
\newblock \bibinfo{title}{{Manganese Oxide/Carbon Aerogel Composite: an
  Outstanding Supercapacitor Electrode Material}}.
\newblock \bibinfo{journal}{Advanced Energy Materials}
  \bibinfo{year}{2011};\bibinfo{volume}{1}(\bibinfo{number}{5}):\bibinfo{pages}{901--907}.
\bibitem[{Fischer et~al.(2007)Fischer, Pettigrew, Rolison, Stroud and
  Long}]{Fischer2007}
\bibinfo{author}{Fischer\xfnm[ A.E.]}, \bibinfo{author}{Pettigrew\xfnm[ K.A.]},
  \bibinfo{author}{Rolison\xfnm[ D.R.]}, \bibinfo{author}{Stroud\xfnm[ R.M.]},
  \bibinfo{author}{Long\xfnm[ J.W.]}.
\newblock \bibinfo{title}{{Incorporation of Homogeneous, Nanoscale MnO2 within
  Ultraporous Carbon Structures via Self-Limiting Electroless Deposition:
  Implications for Electrochemical Capacitors}}.
\newblock \bibinfo{journal}{Nano Letters}
  \bibinfo{year}{2007};\bibinfo{volume}{7}(\bibinfo{number}{2}):\bibinfo{pages}{281--286}.
\bibitem[{Fischer et~al.(2008)Fischer, Saunders, Pettigrew, Rolison and
  Long}]{Fischer2008}
\bibinfo{author}{Fischer\xfnm[ A.E.]}, \bibinfo{author}{Saunders\xfnm[ M.P.]},
  \bibinfo{author}{Pettigrew\xfnm[ K.A.]}, \bibinfo{author}{Rolison\xfnm[
  D.R.]}, \bibinfo{author}{Long\xfnm[ J.W.]}.
\newblock \bibinfo{title}{{Electroless Deposition of Nanoscale MnO2 on
  Ultraporous Carbon Nanoarchitectures: Correlation of Evolving Pore-Solid
  Structure and Electrochemical Performance}}.
\newblock \bibinfo{journal}{Journal of The Electrochemical Society}
  \bibinfo{year}{2008};\bibinfo{volume}{155}(\bibinfo{number}{3}):\bibinfo{pages}{A246--A252}.
\bibitem[{Rouquerol et~al.(1994)Rouquerol, Avnir, Fairbridge, Everett, Haynes,
  Pernicone et~al.}]{IUPAC1984a}
\bibinfo{author}{Rouquerol\xfnm[ J.]}, \bibinfo{author}{Avnir\xfnm[ D.]},
  \bibinfo{author}{Fairbridge\xfnm[ C.W.]}, \bibinfo{author}{Everett\xfnm[
  D.H.]}, \bibinfo{author}{Haynes\xfnm[ J.M.]},
  \bibinfo{author}{Pernicone\xfnm[ N.]}, et~al.
\newblock \bibinfo{title}{{Recommendations for the characterization of porous
  solids (Technical Report)}}.
\newblock \bibinfo{journal}{Pure and Applied Chemistry}
  \bibinfo{year}{1994};\bibinfo{volume}{66}(\bibinfo{number}{8}):\bibinfo{pages}{1739--1758}.
\bibitem[{Wiener et~al.(2004)Wiener, Reichenauer, Scherb and
  Fricke}]{Wiener2004}
\bibinfo{author}{Wiener\xfnm[ M.]}, \bibinfo{author}{Reichenauer\xfnm[ G.]},
  \bibinfo{author}{Scherb\xfnm[ T.]}, \bibinfo{author}{Fricke\xfnm[ J.]}.
\newblock \bibinfo{title}{{Accelerating the synthesis of carbon aerogel
  precursors}}.
\newblock \bibinfo{journal}{Journal of Non-Crystalline Solids}
  \bibinfo{year}{2004};\bibinfo{volume}{350}:\bibinfo{pages}{126--130}.
\bibitem[{Ma et~al.(2006)Ma, Lee, Ahn, Kim, Oh and Kim}]{Ma2006}
\bibinfo{author}{Ma\xfnm[ S.B.]}, \bibinfo{author}{Lee\xfnm[ Y.H.]},
  \bibinfo{author}{Ahn\xfnm[ K.Y.]}, \bibinfo{author}{Kim\xfnm[ C.M.]},
  \bibinfo{author}{Oh\xfnm[ K.H.]}, \bibinfo{author}{Kim\xfnm[ K.B.]}.
\newblock \bibinfo{title}{{Spontaneously Deposited Manganese Oxide on Acetylene
  Black in an Aqueous Potassium Permanganate Solution}}.
\newblock \bibinfo{journal}{Journal of The Electrochemical Society}
  \bibinfo{year}{2006};\bibinfo{volume}{153}(\bibinfo{number}{1}):\bibinfo{pages}{C27--C32}.
\bibitem[{Huang et~al.(2008)Huang, Lv, Yue, Attia and Yang}]{Huang2008}
\bibinfo{author}{Huang\xfnm[ X.]}, \bibinfo{author}{Lv\xfnm[ D.]},
  \bibinfo{author}{Yue\xfnm[ H.]}, \bibinfo{author}{Attia\xfnm[ A.]},
  \bibinfo{author}{Yang\xfnm[ Y.]}.
\newblock \bibinfo{title}{{Controllable synthesis of $\alpha$- and
  $\beta$-MnO(2): cationic effect on hydrothermal crystallization.}}
\newblock \bibinfo{journal}{Nanotechnology}
  \bibinfo{year}{2008};\bibinfo{volume}{19}(\bibinfo{number}{22}):\bibinfo{pages}{225606--225606}.
\bibitem[{Brunauer et~al.(1938)Brunauer, Emmett and Teller}]{Brunauer1938a}
\bibinfo{author}{Brunauer\xfnm[ S.]}, \bibinfo{author}{Emmett\xfnm[ P.H.]},
  \bibinfo{author}{Teller\xfnm[ E.]}.
\newblock \bibinfo{title}{{Adsorption of Gases in Multimolecular Layers}}.
\newblock \bibinfo{journal}{Journal of the American Chemical Society}
  \bibinfo{year}{1938};\bibinfo{volume}{60}(\bibinfo{number}{2}):\bibinfo{pages}{309--315}.
\bibitem[{Lippens and de~Boer(1965)}]{Lippens1965}
\bibinfo{author}{Lippens\xfnm[ B.C.]}, \bibinfo{author}{de~Boer\xfnm[ J.H.]}.
\newblock \bibinfo{title}{{Studies on pore systems in catalysts V. The t
  method}}.
\newblock \bibinfo{journal}{Journal of Catalysis}
  \bibinfo{year}{1965};\bibinfo{volume}{4}(\bibinfo{number}{3}):\bibinfo{pages}{319--323}.
\bibitem[{Harkins and Jura(1944)}]{Harkins1944a}
\bibinfo{author}{Harkins\xfnm[ W.]}, \bibinfo{author}{Jura\xfnm[ G.]}.
\newblock \bibinfo{title}{{Surfaces of Solids. XII. An Absolute Method for the
  Determination of the Area of a Finely Divided Crystalline Solid}}.
\newblock \bibinfo{journal}{Journal of the American Chemical Society}
  \bibinfo{year}{1944};\bibinfo{volume}{66}(\bibinfo{number}{8}):\bibinfo{pages}{1362--1366}.
\bibitem[{Centeno and Stoeckli(2011)}]{Centeno2011}
\bibinfo{author}{Centeno\xfnm[ T.]}, \bibinfo{author}{Stoeckli\xfnm[ F.]}.
\newblock \bibinfo{title}{{Surface-related capacitance of microporous carbons
  in aqueous and organic electrolytes}}.
\newblock \bibinfo{journal}{Electrochimica Acta}
  \bibinfo{year}{2011};\bibinfo{volume}{56}(\bibinfo{number}{21}):\bibinfo{pages}{7334--
  7339}.
\bibitem[{Centeno and Stoeckli(2010)}]{Centeno2010}
\bibinfo{author}{Centeno\xfnm[ T.]}, \bibinfo{author}{Stoeckli\xfnm[ F.]}.
\newblock \bibinfo{title}{{The assessment of surface areas in porous carbons by
  two model-independent techniques, the DR equation and DFT}}.
\newblock \bibinfo{journal}{Carbon}
  \bibinfo{year}{2010};\bibinfo{volume}{48}(\bibinfo{number}{9}):\bibinfo{pages}{2478--2486}.
\bibitem[{Lorrman et~al.(2012)Lorrman, Weber, Reichenauer and
  Pflaum}]{Lorrmann2011}
\bibinfo{author}{Lorrman\xfnm[ V.]}, \bibinfo{author}{Weber\xfnm[ C.]},
  \bibinfo{author}{Reichenauer\xfnm[ G.]}, \bibinfo{author}{Pflaum\xfnm[ J.]}.
\newblock \bibinfo{title}{Electrochemical double-layer charging of
  ultramicroporous synthetic carbons in aqueous electrolytes}.
\newblock \bibinfo{journal}{Electrochimica Acta} \bibinfo{year}{2012};.
\bibitem[{Salitra et~al.(2000)Salitra, Soffer, Eliad, Cohen and
  Aurbach}]{Salitra2000}
\bibinfo{author}{Salitra\xfnm[ G.]}, \bibinfo{author}{Soffer\xfnm[ A.]},
  \bibinfo{author}{Eliad\xfnm[ L.]}, \bibinfo{author}{Cohen\xfnm[ Y.]},
  \bibinfo{author}{Aurbach\xfnm[ D.]}.
\newblock \bibinfo{title}{{Carbon Electrodes for Double-Layer Capacitors I.
  Relations Between Ion and Pore Dimensions}}.
\newblock \bibinfo{journal}{Journal of The Electrochemical Society}
  \bibinfo{year}{2000};\bibinfo{volume}{147}(\bibinfo{number}{7}):\bibinfo{pages}{2486--2493}.
\bibitem[{Tobias and Soffer(1983)}]{Tobias1983}
\bibinfo{author}{Tobias\xfnm[ H.]}, \bibinfo{author}{Soffer\xfnm[ A.]}.
\newblock \bibinfo{title}{{The immersion potential of high surface
  electrodes}}.
\newblock \bibinfo{journal}{Journal of Electroanalytical Chemistry and
  Interfacial Electrochemistry}
  \bibinfo{year}{1983};\bibinfo{volume}{148}(\bibinfo{number}{2}):\bibinfo{pages}{221--232}.
\bibitem[{Eliad et~al.(2001)Eliad, Salitra, Soffer and Aurbach}]{Eliad2001}
\bibinfo{author}{Eliad\xfnm[ L.]}, \bibinfo{author}{Salitra\xfnm[ G.]},
  \bibinfo{author}{Soffer\xfnm[ A.]}, \bibinfo{author}{Aurbach\xfnm[ D.]}.
\newblock \bibinfo{title}{{Ion Sieving Effects in the Electrical Double Layer
  of Porous Carbon Electrodes: Estimating Effective Ion Size in Electrolytic
  Solutions}}.
\newblock \bibinfo{journal}{The Journal of Physical Chemistry B}
  \bibinfo{year}{2001};\bibinfo{volume}{105}(\bibinfo{number}{29}):\bibinfo{pages}{6880--6887}.
\bibitem[{Mysyk et~al.(2009)Mysyk, Raymundo-Pi\~{n}ero and
  B\'{e}guin}]{Mysyk2009}
\bibinfo{author}{Mysyk\xfnm[ R.]}, \bibinfo{author}{Raymundo-Pi\~{n}ero\xfnm[
  E.]}, \bibinfo{author}{B\'{e}guin\xfnm[ F.]}.
\newblock \bibinfo{title}{{Saturation of subnanometer pores in an electric
  double-layer capacitor}}.
\newblock \bibinfo{journal}{Electrochemistry Communications}
  \bibinfo{year}{2009};\bibinfo{volume}{11}(\bibinfo{number}{3}):\bibinfo{pages}{554--556}.
\bibitem[{Kalluri et~al.(2011)Kalluri, Konatham and Striolo}]{Kalluri2011}
\bibinfo{author}{Kalluri\xfnm[ R.K.]}, \bibinfo{author}{Konatham\xfnm[ D.]},
  \bibinfo{author}{Striolo\xfnm[ A.]}.
\newblock \bibinfo{title}{{Aqueous NaCl Solutions within Charged Carbon-Slit
  Pores: Partition Coefficients and Density Distributions from Molecular
  Dynamics Simulations}}.
\newblock \bibinfo{journal}{The Journal of Physical Chemistry C}
  \bibinfo{year}{2011};\bibinfo{volume}{115}(\bibinfo{number}{28}):\bibinfo{pages}{13786--13795}.
\bibitem[{Jiang and Kucernak(2002)}]{Jiang2002}
\bibinfo{author}{Jiang\xfnm[ J.]}, \bibinfo{author}{Kucernak\xfnm[ A.]}.
\newblock \bibinfo{title}{{Electrochemical supercapacitor material based on
  manganese oxide: preparation and characterization}}.
\newblock \bibinfo{journal}{Electrochimica Acta}
  \bibinfo{year}{2002};\bibinfo{volume}{47}(\bibinfo{number}{15}):\bibinfo{pages}{2381--2386}.
\bibitem[{Iwanowski(2004)}]{Iwanowski2004}
\bibinfo{author}{Iwanowski\xfnm[ R.]}.
\newblock \bibinfo{title}{{X-ray photoelectron spectra of zinc-blende MnTe}}.
\newblock \bibinfo{journal}{Chemical Physics Letters}
  \bibinfo{year}{2004};\bibinfo{volume}{387}(\bibinfo{number}{1-3}):\bibinfo{pages}{110--115}.
\bibitem[{Oku et~al.(1996)Oku, Matsuta, Wagatsuma and Konishi}]{Oku1996}
\bibinfo{author}{Oku\xfnm[ M.]}, \bibinfo{author}{Matsuta\xfnm[ H.]},
  \bibinfo{author}{Wagatsuma\xfnm[ K.]}, \bibinfo{author}{Konishi\xfnm[ T.]}.
\newblock \bibinfo{title}{{Simple correlation between Mn-Ka X-ray emission and
  Mn 2p X-ray photoelectron spectra for high oxidation number or low-spin
  manganese compounds}}.
\newblock \bibinfo{journal}{Journal of the Chemical Society, Faraday
  Transactions}
  \bibinfo{year}{1996};\bibinfo{volume}{92}(\bibinfo{number}{15}):\bibinfo{pages}{2759--2764}.
\bibitem[{Madelung et~al.(2000)Madelung, R\"{o}ssler and Schulz}]{DichteMnO2}
\bibinfo{author}{Madelung\xfnm[ M.]}, \bibinfo{author}{R\"{o}ssler\xfnm[ O.]},
  \bibinfo{author}{Schulz\xfnm[ U.]}.
\newblock \bibinfo{title}{{Non-Tetrahedrally Bonded Binary Compounds II}}; vol.
  \bibinfo{volume}{41D} of \emph{\bibinfo{series}{Landolt-B\"{o}rnstein - Group
  III Condensed Matter}}.
\newblock \bibinfo{address}{Berlin/Heidelberg}:
  \bibinfo{publisher}{Springer-Verlag}; \bibinfo{year}{2000}.
\bibitem[{Kanoh et~al.(1997)Kanoh, Tang, Makita and Ooi}]{Kanoh1997}
\bibinfo{author}{Kanoh\xfnm[ H.]}, \bibinfo{author}{Tang\xfnm[ W.]},
  \bibinfo{author}{Makita\xfnm[ Y.]}, \bibinfo{author}{Ooi\xfnm[ K.]}.
\newblock \bibinfo{title}{{Electrochemical Intercalation of Alkali-Metal Ions
  into Birnessite-Type Manganese Oxide in Aqueous Solution}}.
\newblock \bibinfo{journal}{Langmuir}
  \bibinfo{year}{1997};\bibinfo{volume}{13}(\bibinfo{number}{25}):\bibinfo{pages}{6845--6849}.

\end{thebibliography}
\bibliographystyle{model3-num-names}

\end{document}